\documentclass[a4paper,11pt,headings=big,DIV=12]{scrartcl}
\usepackage{amsmath,amssymb}
\usepackage{color,xcolor}
\definecolor{blue}{rgb}{0,0,0.5} 
\usepackage[colorlinks,linkcolor=blue,urlcolor=blue,citecolor=blue]{hyperref}
\usepackage{graphicx}
\addtokomafont{disposition}{\rmfamily\boldmath}
\usepackage[utf8]{inputenc}
\usepackage{enumerate} 
\usepackage{slashed}
\usepackage{cite} 
\usepackage{comment}
\usepackage{multirow}
 \usepackage{bbold}
\allowdisplaybreaks
\newcommand{\eom}{\text{eom}}

\newcommand{\ba}{\be_a}
\newcommand{\bb}{\be_b}
\newcommand{\bc}{\be_c}
\newcommand{\barb}{\bar{b}}

\newcommand{\ONE}{\mathbb{1}}

\newcommand{\zamD}[1]{\chi_{#1#1}}
\newcommand{\zam}[2]{\chi_{#1 #2}}

\newcommand{\zamDR}[2]{\chi_{#1#1}^{#2}}
\newcommand{\zamR}[3]{\chi_{#1 #2}^{#3}}

\newcommand{\CW}{{\mathbb C}}
\newcommand{\fin}{[\text{finite}]}

\newcommand{\MS}{{ \textrm{MS}}}

\newcommand{\plnas}{ \partial_{\ln \als}}
\newcommand{\pas}{ \partial_{ \als}}
\newcommand{\als}{{a_s}}

\newcommand{\alsUV}{{a_s^\UV}}

\newcommand{\alsp}[1]{{a_s^{#1}}}
\newcommand{\gaZ}[1]{{\ga}_{#1,0}}

\newcommand{\CWI}[3]{\CW_{#1 #2}^{#3}}
\newcommand{\CWID}[2]{\CW_{#1 #1}^{#2}}
\newcommand{\cwi}[2]{\CW_{#1 #2}^{\ONE}}
\newcommand{\cwiD}[1]{\CW_{#1 #1 }^{\ONE}}

\newcommand{\cwiR}[3]{\CW_{#1 #2}^{\ONE, #3 }}
\newcommand{\cwiDR}[2]{\CW_{#1 #1}^{\ONE, #2}}

\newcommand{\lDfR}[2]{  \bar{L}_{#1 #1}^{\ONE,#2}}

\newcommand{\LDR}[3]{L_{#1 #1}^{#2,#3}}
\newcommand{\LnDR}[4]{L_{#1 #2}^{#3,#4}}

\newcommand{\lDR}[2]{ L_{#1 #1}^{\ONE,#2}}

\newcommand{\lnDR}[3]{ L_{#1 #2}^{\ONE,#3}}

\newcommand{\lD}[1]{ L_{#1 #1}^{\ONE}}

\newcommand{\rnDI}[3]{r_{#1 #2 }^{\ONE (#3)}}
\newcommand{\rDI}[2]{r_{#1 #1 }^{\ONE (#2)}}

\newcommand{\lnD}[2]{ L_{#1 #2}^{\ONE}}

\newcommand{\gc}[1]{{g^{#1}}}
\newcommand{\gcb}[1]{{g_0^{#1}}}

\newcommand{\ZopI}[2]{{ \Zop_{#1}^{\phantom{#1} #2}}  }
\newcommand{\ZopII}[2]{{ (\Zop^{-1})_{#1}^{\phantom{#1} #2}}  }

\newcommand{\Zop}{{\mathbb{Z}}}

\newcommand{\Bone}[1]{{b_{1,#1}}}

\newcommand{\TEMT}[1]{ T^{#1}_{ \;\;#1}}
\newcommand{\TEMTO}{\Theta}
\newcommand{\TEMTG}{\Theta_{\textrm{gravity}}}
\newcommand{\TEMTE}{\Theta_{\textrm{eom}}}
\newcommand{\TEMTgf}{\Theta_{\textrm{gf}}}
\newcommand{\gsym}{g}

\newcommand{\Dloc}[2]{{  \de_{ #1 (#2)}}}

\newcommand{\Dglob}[1]{{  \partial_{ #1}}}

\newcommand{\VEV}[1]{\left\langle #1 \right\rangle} 
\newcommand{\vev}[1]{\langle #1 \rangle}

\newcommand{\matel}[3]{\langle #1|#2|#3\rangle}
\newcommand{\al}{\alpha}
\newcommand{\be}{\beta}
\newcommand{\ga}{\gamma}
\newcommand{\de}{\delta}

\newcommand{\eps}{\epsilon}

\newcommand{\Rtr}[2]{\Theta}

\newcommand{\UV}{{\textrm{UV}}}
\newcommand{\IR}{{\textrm{IR}}}

\newcommand{\LUV}{\Lambda_{\UV}}

\newcommand{\tr}{{\textrm{ tr}}}

\newcommand{\Zpart}{{\cal Z}}

\newcommand{\Og}{O_{g}}

\newcommand{\Ra}{  { \RR_1} }
\newcommand{\RR}{  { {\cal R}} }
\newcommand{\Rb}{  { \RR_2} }

\newcommand*{\mathcolor}{}
\def\mathcolor#1#{\mathcoloraux{#1}}
\newcommand*{\mathcoloraux}[3]{%
  \protect\leavevmode
  \begingroup
    \color#1{#2}#3%
  \endgroup
}

\begin{document}

\begin{flushright}
\begin{tabular}{l}
CP3-Origins-2016-004 DNRF90 \\
 DIAS-2016-4
\end{tabular}
\end{flushright}
\vskip1.5cm

\begin{center}
{\Large\bfseries \boldmath On Finiteness of $2$- and $3$-point Functions and \\[0.1cm] the Renormalisation Group }\\[0.8 cm]
{\Large%
Vladimir Prochazka$^{a,b}$
and Roman Zwicky$^a$,
\\[0.5 cm]
\small
$^a$ Higgs Centre for Theoretical Physics, School of Physics and Astronomy,\\
University of Edinburgh, Edinburgh EH9 3JZ, Scotland  \\[0.2cm]
$^b$ Weizmann Institute of Science, Rehovot, 76100, Israel
} \\[0.5 cm]
\small
E-Mail:
\texttt{\href{mailto:v.prochazka@ed.ac.uk}{v.prochazka@ed.ac.uk}},
\texttt{\href{mailto:roman.zwicky@ed.ac.uk}{roman.zwicky@ed.ac.uk}}.
\end{center}

\bigskip
\pagestyle{empty}

\begin{abstract}
Two and three point functions of composite operators are analysed   
with regard to (logarithmically) divergent contact terms. 
Using the renormalisation group  of  dimensional regularisation it is established that 
the divergences are governed by the anomalous dimensions of the operators and 
the leading UV-behaviour of the  $1/\eps$-coefficient.    
 Explicit examples are given by the $\vev{G^2G^2}$-,  $\vev{\TEMTO \TEMTO}$-
(trace of the energy momentum tensor)  and  $\vev{\bar q q \bar q q}$-correlators in 
QCD-like theories. The former two are  convergent when the $1/\eps$-poles are resummed 
but divergent at fixed order  implying that  perturbation theory and  
 the $\eps \to 0$ limit do not generally commute.  Finite correlation functions obey 
 unsubtracted dispersion relations which is of importance when they are 
 directly related to  physical observables.
As a byproduct the $R^2$-term of the trace anomaly is extended to NNLO (${\cal O}(\alsp{5})$), in the $\MS$-scheme, using a recent $\vev{G^2G^2}$-computation. 
\end{abstract}

\newpage

\setcounter{tocdepth}{3}
\setcounter{page}{1}
\tableofcontents
\pagestyle{plain}

\section{Introduction}
\label{sec:intro}
 
In this paper  divergences are investigated which arise when  (composite) operators approach each 
other. 
These ultraviolet (UV) divergences are necessarily local  and expressed in terms of delta functions and 
derivatives thereof (i.e. contact terms (CTs)). This requires renormalisation in addition to the parameters
of the theory and the composite operators themselves. 
 
 These CTs play an important role as they manifest themselves as anomalies in 
 correlation functions of composite operators, 
the chiral anomaly serving as a primary example\footnote{Early analyses 
centred around  configuration space singularities in correlation functions, 
without particular emphasis on perturbation theory, of the chiral and trace anomalies 
can be found in \cite{Wilson:1969zs,C72} and \cite{C72,CE72,CE72b} and 
reviewed in \cite{Crewther} respectively. 
Recently CTs in $3$-point correlation functions were the centre of discussion 
on whether  in $d=4$ non-trivial unitary scale but not  conformal field theories exist \cite{KL14,CommentsSCFT}.}
   and the perspective on other anomalies 
continues to evolve \cite{anomaly1,anomaly2,Skenderis}. On the other hand CTs are  
 not important when studying 
the  spectrum of $2$-point functions (e.g. QCD sum rules \cite{SVZ}) or lattice QCD \cite{DeGrand:2006zz})
since they bear no relation to the infrared (IR) spectrum. In lattice simulations 
of correlation functions CTs require additional renormalisation conditions, a problem 
for which the $4+1$-dimensional gradient flow offers new perspectives \cite{NN06,L10,LPoS}.

Our work originates from the observation that the leading logarithm (LL) $\epsilon$-poles 
of the field strength tensor 
correlation function 
 sums to an expression  
\begin{eqnarray}
\label{eq:LgLL0}
\int d^4 x e^{i x \cdot p} \vev{[G^2(x)] [G^2(0)]} |_{\textrm{LL}-\text{poles}}  \;   &\;\sim \;&  \;  p^4  \frac{1}{\epsilon + \be_0 \als}      \nonumber \\[0.1cm]
 &\;=\;&  p^4 \frac{1}{\epsilon} \left( 1 - \frac{\be_0 \als}{\epsilon} +\frac{(\be_0 \als)^2}{\epsilon^2}    + {\cal O}(\als^3)  \right)  \;,
\end{eqnarray}
which is finite for $\epsilon \to 0$ but divergent at each fixed order in perturbation theory.
Using the renormalisation group (RG) of dimensional regularisation (DR) the absence of potential 
logarithmic divergences  is systematised 
in various ways.
Firstly, simple criteria for convergence, involving RG-quantities,
are established  of generic $2$-point functions. 
The discussion is extended to include the non-perturbative  condensate terms, mutiple couplings 
and $3$-point functions.  Using the local quantum action principle (QAP) 
a closed integral expression for the  $R^2$-anomaly is given in terms the the first pole of the 
the correlation function \eqref{eq:LgLL0}.

The paper is organised as follows. In section \ref{sec:finite2}
the finiteness criteria for $2$-point functions 
are discussed, followed by the explicit examples of $\vev{G^2G^2}$- and $\vev{\TEMTO \TEMTO}$-correlators 
in QCD-like theories in section \ref{sec:QCD-like}.  Implications for dispersion integrals, 
RG-scale dependence (physicality) and 
the $R^2$-anomaly are elaborated on in sections  \ref{sec:disp}, \ref{sec:physical} and  \ref{app:R2-anomaly} respectively. 
The $1$-coupling case of the $2$-point function is generalised
to  multiple couplings and $3$-point functions in sections  \ref{sec:multiple}  and 
 \ref{sec:3ptMom}.
 The paper ends with 
a summary and conclusions in section \ref{sec:conclusions}.
Appendix \ref{app:A} contains details about the $\vev{G^2 G^2}$-correlation function computation and 
appendix \ref{app:quark} discusses  the convergence of the $\vev{\bar q q \bar q q} $- and
$\vev{J_\mu^5J_\nu^5 }$-correlation functions.
 The $\be$-function conventions are given in appendix \ref{app:beta}.

\section{$2$-point Function in Momentum Space}
\label{sec:finite2}

We consider a renormalisable theory (i.e. with a UV fixed point (FP)) in four dimensions 
with no explicit mass scales and a non-trivial flow.
 The euclidean $2$-point functions of marginal operators are parametrised as follows\footnote{Various extensions of this set up will be discussed: 
condensate corrections in the language of the operator product expansion (OPE)  \cite{Wilson:1969zs}, non-diagonal correlation functions, 
multiple couplings and 3-point functions are discussed in sections 
\ref{sec:OPE}, \ref{sec:multiple} and \ref{sec:3ptMom} respectively. 
An extension to operators with spin is possible and we refer the reader to \cite{Boch1} where this is done in the context of  QCD for diagonal correlation functions.}\,
\footnote{The correlation function $\vev{[O_A(x)][ O_B(0)]}_c$ is formally defined 
by the connected part of a regularised path-integral representation $\int D \phi [O_A(x)][ O_B(0)] e^{-S[\phi]}$.}
\begin{eqnarray}
\label{eq:GAB}
\Gamma_{AB}(p^2)   =   
 \int d^4 x  e^{i p \cdot x} \vev{[O_A(x)][ O_B(0)]}_c   =
 \cwi{A}{B} (p^2)  p^4  \;,
\end{eqnarray}
where $c$ stands for the connected component, $\vev{\dots}$ for 
the vacuum expectation value (VEV),  
$[O_{A,B}]$ are renormalised scalar (composite) operators
   of mass dimension $4$  
and $\cwi{A}{B}$  are dimensionless functions. 
Such a divergence  might be thought of as the Wilson coefficient of the identity operator. 
In an asymptotically free (AF) theory the coefficient $\cwi{A}{B} (p^2)$ is potentially logarithmically divergent by power counting. 
In coordinate space this divergence results from  singular behaviour as $x \to 0$. 
The latter can be removed by local counterterms 
 within the standard 
renormalisation programme.  The renormalised correlation function  
$\Gamma_{AB}^{\RR}$
is obtained from the bare one $\Gamma_{AB}$ 
by splitting the bare Wilson coefficient 
 $\cwi{A}{B} (p^2) $ into  renormalised  $\cwiR{A}{B}{\RR} (p^2)$ and a counterterm 
 $\lnDR{A}{B}{\RR} $ part 
\begin{equation}
\label{eq:subtract}
  \cwi{A}{B} (p^2) =  \cwiR{A}{B}{\RR} (p^2) +  \lnDR{A}{B}{\RR} \;.
\end{equation} 
Above, the letter  $L$ either stands  for local and R denotes a renormalisation scheme. 
To be clear we wish to add that $ \cwiR{A}{B}{\RR}$ is finite whereas  $\lnDR{A}{B}{\RR}$ is generally not despite the R-label. We are going to be careful as to which statements are generic for any scheme (i.e. a specific split in \eqref{eq:subtract}) and what is valid when $\RR$ stands for 
the minimal subtraction ($\MS$) scheme.

In coordinate space this translates into
\begin{equation}
\label{eq:xspace}
\hat{\Gamma}_{AB}^{\RR} (x^2)  =  \underbrace{ \vev{[O_A(x)][ O_B(0)]}_c}_{\equiv \hat{\Gamma}_{AB} (x^2)}     - \lnDR{A}{B}{\RR}
 \Box^2 \delta(x) \;,
  \end{equation}
where $\Box = \partial_\mu \partial^\mu$ and $\de(x)$ is the four dimensional Dirac delta function throughout. With slight abuse in notation we refer to $\cwi{A}{B} (p^2)$ as the bare correlation
function despite its dependence 
on the RG-scale through the renormalisation of he composite operators $[O_{A,B}]$.
The  renormalised correlation function 
$\cwiR{A}{B}{\RR} (p^2)$ and the counterterm  $\lnDR{A}{B}{\RR}$ are in general RG-scale 
dependent even if $[O_{A,B}]$ are not.  This particular RG-scale dependence of course cancels in the sum and consists of CTs since $\lnDR{A}{B}{\RR}$ is local.

\subsection{$2$-point Function in Dimensional Regularisation with one Coupling}
\label{sec:General2pt}

At first we restrict ourselves to one coupling 
$\als = \als(\mu)$ whose scale dependence will frequently be suppressed throughout.  
In the $\MS$-scheme with DR ($d = 4- 2 \eps$)  
the  counterterm 
\begin{equation}
 \lnDR{Q}{Q}{\MS}(\mu)  \equiv  \sum_{n \geq 1} \frac{\rnDI{Q}{Q}{n} (\als(\mu))}{\eps^n}   \;, \qquad 
\rnDI{Q}{Q}{1}(\als) = \rnDI{Q}{Q}{1,0} + \rnDI{Q}{Q}{1,1} \, \als + {\cal O}(\alsp{2})  \;,
\end{equation}
is given by 
 a Laurent series. 
 The residues $\rnDI{Q}{Q}{n}$ are dimensionless  and functions of the  running coupling only. 
 Since in this work we use the $\MS$-scheme in all practical computation we do 
 not indicate this circumstance with a further label.
We proceed to derive a RG-equation (RGE) for  $\lnD{Q}{Q}$. The starting point is 
\eqref{eq:subtract}, which  in DR  
\begin{equation}
\label{eq:QQfin}
 \cwiD{Q} (p^2)  p^{-2 \eps}   =   
  \big( \cwiDR{Q}{\RR} (p^2)   +  \lnDR{Q}{Q}{\RR}  \big) \mu^{-2 \eps}   \;.
\end{equation}
The renormalised Wilson coefficient   $ \cwiDR{Q}{\RR} (p^2)$ is finite 
for $\eps \to 0$ in the sense of being analytic in  $\eps$ (in particular no poles).
Suppose that  $ [O_Q]$  can be made RG-invariant by premultiplying by a finite factor $\kappa_Q(\als)$, i.e. 
\begin{equation}
\label{eq:IRG}
 \frac{d}{d \ln \mu} \kappa_Q  Z_{QQ} = 0   \;, \quad Z_{QQ} O_Q =  [O_Q] + \dots \;,
 \end{equation}
where the dots correspond to equation of motion (eom) operators which do not contribute to structure we are discussing.  
  Using  
  $ \frac{d}{d \ln \mu} \kappa_Q^2 \cwiD{Q} (p^2)  = 0$  and the finiteness of the renormalised Wilson coefficient one deduces\footnote{In the case where $O_Q$ is marginal (and not $m \bar q q$)   $\kappa_Q = \hat{\be}^Q$ and $\hat{\ga}_Q = 2 \partial_{\ln \als} \hat{\be}^Q$, the non $\eps$-part becomes Lie derivative,
$ {\cal L}_{\be} = 2  \hat{\ga}_Q  +   \hat{\be}^P \partial_P  $, acting on 
the two tensor $\lDR{Q}{\RR}$. This circumstance is put into evidence in the multiple coupling section \ref{sec:extension} which reveals the structure more systematically.}
    \begin{equation}
\label{eq:RGE1}
(2 \hat{ \ga}_Q  +   \frac{d}{d \ln \mu} - 2\eps)     \lDR{Q}{\RR} =  - \zamDR{Q}{\RR} 
 \;, \quad 
\end{equation}
where 
\begin{equation}
\label{eq:ad}
\hat{ \ga}_Q =  \frac{d}{d \ln \mu} \ln \kappa_Q   \;,
 \end{equation}
with  $\hat{\be} = - \eps + \be$ and $\hat{\ga}_Q = \ga_Q - \xi_Q \eps$ 
 being the $d$-dimensional  $\be$-function and anomalous dimension respectively.\footnote{Note that the  $\ga_Q$'s refer to the anomalous dimensions of the operators and not to the $\kappa_Q$-parameters ($ \ga_Q = - \ga_{\kappa_Q}$). Using that for a $1$ coupling theory 
 and a mass independent scheme  $\frac{d}{d \ln \mu}  = 2 \hat{\be} \plnas $, \eqref{eq:RGE1} can be written as $\left( ( \eps -    \hat{\ga}_Q) - \hat{\be} \plnas  \right) \lDR{Q}{\RR}  =   \zamDR{Q}{\RR}/2 $.  }
 The quantity $ \zamDR{Q}{\RR}$ follows from the requirement of finiteness and is given 
 in the $\MS$-scheme by 
 \begin{equation}
\label{eq:zamMS}
 \zamDR{Q}{\MS} = 2( \als  \pas ( \als \rDI{Q}{1} )  + \xi_Q \rDI{Q}{1}) \;.
\end{equation}
The ordinary differential equation \eqref{eq:RGE1} is solved by 
\begin{equation}
\label{eq:LQu}
\lDR{Q}{\RR}(\mu) =    \int_{\ln \mu}^{\infty} \zamDR{Q}{\RR} ( \als(\mu') )   I_{{QQ}}( \mu ,\mu')
 \left( \frac{\mu}{\mu'} \right)^{2 \eps} d \ln \mu'  \;,
\end{equation}
which shows the $\MS$-property  that all higher pole residues  of $\lD{Q}$   follow from the first one 
(encoded in $\zamD{Q}$ \eqref{eq:zamMS}).
Above   $ (\mu/\mu')^{2 \eps}  I_{{QQ}}$  is an integrating factor with
\begin{eqnarray}
\label{eq:IF}
I_{{QQ}}(\mu,\mu')  =  \exp \left( 2  \int_{\ln \mu}^{\ln \mu'}   \hat{\ga}_Q( \als(\mu'')) 
d \ln \mu'' \right)   \;.
\end{eqnarray}
Generally it is the function $I_{QQ}$ and the power behaviour of 
$\zamD{Q}$  which decide on whether or not the integral diverges for $\mu' \to \infty$
and   $(\mu/\mu')^{2 \eps} $ serves as  a potential UV-regulator. 
A more refined analysis is required to distinguish whether the UV-FP is 
of the AF-type $\alsUV \equiv \als(\infty) = 0$ or asymptotically safe (AS)-type $\alsUV \neq 0$.

\subsubsection{Asymptotically free Theory}
\label{sec:AF}

For the asymptotic analysis it is convenient to  change the variable to the RG-time  
$ t \equiv \ln \mu/\mu_0$. 
In the asymptotic regime a LL analysis  is sufficient.   
Assuming $\hat{\be}(\als) = - \eps+  \be = - \eps - \be_0 \als + {\cal O}(\alsp{2})$  
the LL relation is given by\footnote{For 
$\hat{\be}(\als) = - \eps+  \be = - \eps - \be_0 \alsp{r} + {\cal O}(\alsp{r+1})$ this leads to 
$\als(t) = ( \als(\mu) e^{-2 \eps t} \eps^{1/r} ) / ( \eps + \be_0 \alsp{r}(\mu) (1 - e^{-2 r \eps t }) )^{1/r} \to
 \als(\mu) ( 1+ 2r \be_0 \alsp{r}(\mu) t)^{-1/r}$ for $\eps \to 0$ provided $r > 0$. 
  In the case where $\ga_Q(\als) \sim \alsp{r}$ the formula \eqref{eq:convAF} still applies. For more generic cases we leave it to the reader to work out the relevant formula from 
  Eqs.~(\ref{eq:LQu},\ref{eq:IF}).}
\begin{equation}
\als(t) = \frac{\eps \als(\mu) e^{- 2 \eps t}}{\eps + \be_0 \als(\mu) (1- e^{-2 \eps t})} = \frac{\als(\mu)}{1+  2 \be_0 \als(\mu)  t }  + {\cal O}(\eps)  \;,
\end{equation}
with slight abuse in notation and initial value $\als(t=0) = \als(\mu) $ (for $\eps \to 0$)  and UV-value $\als(t \to \infty ) =0$.
The anomalous dimension is parameterised by $\ga_Q = \als \gaZ{Q} +  {\cal O}( \alsp{2})$  implying the asymptotic behaviour $ I_{{QQ}}(t) \sim t^{\eta}  $ with 
$ \eta =  \gaZ{Q}/\be_0$.
 Assuming  
 $\zamD{Q} \sim t^{-n_{QQ}}$ to be perturbative for $ t \to \infty$ with $n_{\textrm{QQ} }\geq 0$
($n_{QQ}=0$, i.e. $\zamD{Q} = {\cal O}( \alsp{0})$, being the nominal case for a non-trivial unitary theory) 
the condition for  UV finiteness is\footnote{In the case of a non-diagonal correlation function, 
as in \eqref{eq:GAB}, 
with a single coupling theory the criterion \eqref{eq:convAF} generalises to 
$1+  \frac{\gaZ{A}+\gaZ{B}}{2 \be_0}  < n_{AB}$ where the operator basis has been 
assumed to be diagonalised at LO. 
An example is given by QCD with the topological term $O_1 =  G \tilde G$ and 
$O_2 = \partial_\mu \bar q \ga^\mu \ga_5 q$ which do mix with each other 
$Z_{1}^{\phantom{1}2} = 12 C_F \als \frac{1}{\eps} + {\cal O }( \alsp{2})$.}
\begin{equation}
\label{eq:convAF}
1+  \frac{\gaZ{Q}}{\be_0}  < n_{QQ}  \quad   \Leftrightarrow \quad 
  \lDR{Q}{\RR}   \stackrel{\epsilon/(\be_0 \als)  \to 0}{\longrightarrow}   \lDfR{Q}{\RR}    = \fin  \;.
 \end{equation} 
 This result is presumably scheme-independent since, as it is well-known, both $\be_0$
 and $\gaZ{Q}$ are scheme-independent.
 
 \paragraph{Leading behaviour in the $\MS$-scheme} The leading behaviour of $\lDR{Q}{\MS}$ is obtained explicitly by using the $1-$loop expressions for $\beta, \gamma_Q$ and the LO of $\zamDR{Q}{\MS}$ 
 involves hypergeometric functions and is given in \eqref{eq:hyper} in appendix \ref{app:detail}.
For $\xi=0$ \eqref{eq:hyper} becomes simpler 
 which can be expanded in $\als$
\begin{eqnarray}
\label{eq:LQresult}
 \lDR{Q}{\MS}(\mu)  &\simeq&  2  \rDI{Q}{1,0}  \int_{0}^{\infty}  e^{-2 \eps t} \eps^{-\frac{\gaZ{Q}}{\be_0}}  (\eps+\be_0 \als (1-e^{-2\eps t})^{\frac{\gaZ{Q}}{\be_0}}) dt 
\nonumber  \\[0.1cm]
 &=& \rDI{Q}{1,0} \frac{(1+ \frac{\als \be_0}{\eps})^{1+\frac{\gaZ{Q}}{\be_0}}-1}{ \als (\be_0+\gaZ{Q})} \notag \\ 
&=& \rDI{Q}{1,0}  \left( \frac{1}{\eps}+\frac{\gaZ{Q} \als}{2 \eps^2}+\frac{(-\be_0
\gaZ{Q}+(\gaZ{Q})^2)\alsp{2}}{6  \eps^3} + {\cal O}(\alsp{3})  \right) \;,
\end{eqnarray} 
where the $\mu$-dependence  arises from  $\als = \als(\mu)$.
There are  divergent terms at each order in the $\als$-expansion.
Provided \eqref{eq:convAF} is met for $n_{QQ}=0$ the $\eps \to 0$ limit is finite and gives
\begin{equation}
\lDfR{Q}{\MS}  =   - \frac{\rDI{Q}{1,0}}{ \als (\be_0 + \gaZ{Q})}  \;.
\end{equation}

Two important remarks are in order. First when \eqref{eq:LQresult} is expanded 
in powers of $\als$ then $1/\eps$-poles appear irrespective of whether condition 
\eqref{eq:convAF} is obeyed or not. 
This is an example of where fixed order perturbation theory gives 
the wrong indication about convergence. 
Secondly, even though convergent 
 the $ \eps \to 0$  followed by $\als \to 0$ limit  
does not exist for the $\vev{O_QO_Q}$-correlation function. 
That is to say that in general  the $\als$-expansion (fixed order)  and $\eps \to 0$-limit do 
\emph{not} commute.  In the cases 
where the correlation function is related to a physical observable, such as the  
trace of the energy moment tensor (TEMT) correlation function, 
there are $\als$-dependent prefactors which ensure a smooth limit. 

\subsubsection{Asymptotically  safe Theory} 
\label{sec:AS}

The non-trivial FP is characterised by generally non-vanishing 
anomalous dimensions
$\ga_Q =  \ga_Q^* +  (\als- \alsUV) \gaZ{Q} + \dots$. The integrating factor assumes the form 
$ I_{{OQ}}(t)   \sim e^{ 2 \ga_Q^*t} $.  The exponential behaviour dominates over the  polynomial 
behaviour of $\zamDR{Q}{\RR} $.     
Hence the sign of $ \ga_Q^*$ determines the convergence
\begin{equation}
\label{eq:convAS}
 \ga_Q^*   < 0   \quad \Rightarrow \quad   \lDR{Q}{\RR}   \stackrel{\epsilon/\als \to 0}{\longrightarrow}   \fin    
 \;.
\end{equation}
If $ \ga_Q^*  > 0$ , $\lDR{Q}{\RR}$ diverges and if $ \ga_Q = 0$ then the analysis of AF-case in the previous 
section applies.

\subsection{Summary and Contemplation}

In summary the presence or absence of  UV divergences 
depends on the anomalous dimension $\ga_Q$ and the leading power behaviour of the 
$\zamD{Q}$. A more detailed comparison is instructive.
In the AF-case \eqref{eq:convAF} the condition depends on both quantities mentioned 
above whereas
in the AS-case \eqref{eq:convAS} it only depends on the anomalous dimension at the FP. The polynomial behaviour of $\zamD{Q} $ is overruled by the exponential behaviour of the anomalous dimension. This is reminiscent of marginal flows requiring specific analysis
in order to determine whether or not they are relevant or exactly marginal,  
 whereas relevant and irrelevant flows are settled from the start. 
The behaviour of the AS-case is similar to the case of a scale  or conformaly invariant 
field theory. 
The $2$-point function of  operators, of scaling dimension $\Delta_O = d_O + \ga_O$, is given by 
$ \vev{O(x)O(0)} \sim  {(x^2)^{-\Delta_O} }$.
In our case $d_O = 4$ and the Fourier transform of the $p^4$-structure is convergent provided 
$\ga_O < 0$ in accordance with the criteria for an AS theory \eqref{eq:convAS}.

A priori the divergent structure of $2$-point function  of dimension four 
operators in momentum space reads ($d=4$)
\begin{equation}
\label{eq:GABdiv}
\Gamma_{AB} \sim a \,\LUV^4 + b\,  p^2  \LUV^2 + c \, p^4 \ln \LUV + \fin  \;,
\end{equation}
for a cut-off regularisation. 
Above $a,b,c$ are dimensionless functions of $\LUV/\mu_0$ where $\mu_0$ is some reference scale.
In this section it was shown  under what conditions $c_{\textrm{DR}}( \LUV/\mu_0)   \ln \LUV  =  \fin$ holds for $\LUV \to \infty$ in DR (symbolically $ \ln \LUV  \leftrightarrow 1/\eps$).  Since DR is defined only in perturbation theory 
one might  question as to whether the result holds 
outside this framework.  
An argument in favour is that perturbation theory is trustworthy in the UV and that the 
LL approximation should  therefore be sufficient. One assumption  though is that the UV 
divergences can be captured as a Laurent expansion in powers of $1/\eps$. 
Whether or not this is valid outside perturbation theory is unknown since DR 
is only defined perturbatively. 
It is well-known that DR is blind to power divergences since no explicit scale is introduced 
into the integral regularisation other than the prefactor $\mu^{- 2 \eps}$. 
Hence $a_{\textrm{DR}} = b_{\textrm{DR} }=0$ is built into DR rather than being 
a result.\footnote{Let us mention, in passing, that it has been argued by Bardeen  \cite{Bardeen:1995kv} 
 that  cut-off 
 regularisations  are not a natural choice for renormalisable theories.
 For example when a theory exhibits a global chiral symmetry one would preferably  
 use a chirally invariant regularisation as otherwise the Ward Identities need to be fixed by 
 adding local counterterms.} 
 
In sections \ref{sec:multiple}, \ref{sec:3ptMom} the results are  generalised straightforwardly to the case of multiple couplings  and  3-point functions. 
First, we illustrate the findings in the familiar setting of QCD-like gauge theories including an extension of the $R^2$-anomaly to one order higher.
This will clarify the meaning of the  quantity $\zamD{Q}$ as being related to trace anomaly of the external sources of the corresponding operators cf. \cite{JO90}.

\section{QCD-like Gauge Theory as an Example} 
 \label{sec:QCD-like}
 
We consider a  QCD-like gauge theory, i.e. $N_f$ massless fermions in a fundamental representation coupled to gluons in the adjoint representation for a $SU(N_c)$ gauge group. 
This implies in particular a non-trivial RG-flow.
In section \ref{sec:gluon} finiteness of 
the $\vev{G^2 G^2}$- and the closely related $\vev{\TEMTO \TEMTO}$-correlators is 
established,\footnote{The $\vev{ \bar q q \bar q q}$ and $\vev{J_\mu^5J_\nu^5 }$-correlation functions 
are discussed in appendices \ref{app:mass} and \ref{app:55} respectively.}
followed by a discussion of the physical consequences:  
unsubtracted dispersion relation (section \ref{sec:disp}) and observability of the bare correlation function (section \ref{sec:physical}).
In section \ref{sec:OPE} the discussion is extended to include  condensates  through the OPE.

\subsection{Gauge Theory Correlation Functions}
\label{sec:gluon}

\paragraph{Correlation functions of the field strength tensor}

We consider the  $2$-point function of the field strength correlation function squared, with   
\begin{equation}
\label{eq:Og}
[O_g] = [\frac{1}{g_0^2} G^2]   \;,
\end{equation}
where $G^2 = G_{\mu \nu}^2$ is 
is the usual field strength tensor $G_{\mu\nu} = - i [ D_\mu ,D_\nu]$ squared with covariant derivative $D_\mu = (\partial + i A)_\mu$.
From \eqref{eq:IRG}   $ \kappa_g = \hat{\be} $   and therefore 
$\ga_g = 2   \plnas   \hat{\be}$ follows. This leads to   a simple form of the integrating factor  \eqref{eq:IF}
\begin{equation}
 I_{gg}(\mu,\mu')  =  \left(  \frac{\hat{\be}(\mu')}{\hat{\be}(\mu)} \right)^2 \;.
\end{equation} 
The corresponding Laurent series \eqref{eq:LQu}, changing variables 
to $d \ln \mu' = d u / (2u \hat{\be}(u))   $  takes on the form  
\begin{equation}
\label{eq:Lgeps}
\lDR{g}{\RR}(\mu,\eps) =  - \frac{1}{2 \hat{\be}^2(\als(\mu))}  \int_0^{\als(\mu)} \zamDR{g}{\RR}(u) \hat{\be}(u) \left( \frac{\mu(\als)}{\mu(u)} \right)^{2 \eps}
 \frac{du}{u}   \;,
\end{equation}
where the factor $(\dots)^{2 \eps}$ will be specified further below.
This expression is convergent as  $\gaZ{g} = - 2 \be_0$ and $\zamDR{g}{\RR}(\als)  \sim {\cal O}(\alsp{0})$ obey 
the inequality  \eqref{eq:convAF} with $1- 2 < 0$.  This means that 
\begin{equation}
 \lDR{g}{\RR}(\eps)   \stackrel{\epsilon/ \be  \to 0}{\longrightarrow}   \lDfR{g}{\RR}  =  \fin \;.
 \end{equation}
It is instructive to consider this constant explicitly at LL  in the $\MS$-scheme
\begin{equation}
\label{eq:LgLL}
\lDR{g}{\MS}|_{\textrm{LL}}  =  \frac{\rnDI{g}{g}{1,0}}{\epsilon + \be_0 \als}    \stackrel{\epsilon/ (\be_0 \als) \to 0}{\longrightarrow}   \lDfR{g}{\MS}|_{\textrm{LL}} \equiv  \frac{\rnDI{g}{g}{1,0}}{\be_0 \als}  = \fin \;,
\end{equation}
as it becomes apparent that the correlation function is not finite for  $\als \to 0$.  
One should keep in mind that the field strength correlation function is not a physical quantity unlike the closely 
related correlation function of the TEMT discussed below.
Before doing so let us emphasise that when expanding \eqref{eq:Lgeps} in $\als$ divergent terms appear cf. \eqref{eq:LQresult}. 
We indeed reproduce the divergent terms in  \cite{Kataev:1981gr, Zoller:2012qv} at 
NLO (next leading order) and NNLO respectively. 
 To obtain agreement it is important to expand the term to the power $2 \eps$ in the integrand $ ( \mu(\als)/\mu(u))^{2 \eps} = \exp  ( \eps \int_{u}^{\als} du'/ (u' \hat{\be}(u')))=  u/\als(\mu) +  {\cal O}(1/\eps)$.

\paragraph{Correlation functions of the trace of the energy momentum tensor}

The TEMT decomposes as follows 
\begin{equation}  
\label{eq:TEMT}
\vev{\TEMT{\rho}} =  (-\de_{s(x)}) \ln \Zpart =  \vev{ \TEMTO } +  \vev{\TEMTG }  + \vev{\TEMTE }  + \vev{\TEMTgf  } \;, \quad     \TEMTO  =  \frac{\hat{\be}}{2}  [O_g] \;,
\end{equation}  
where $\TEMTO,\TEMTG, \TEMTE,\TEMTgf   $ are the operator,  curvature dependent,  \eom\; and gauge fixing  part of the TEMT. The  $\TEMTgf$-part does not contribute to 
physical observables, $\TEMTG$ vanishes in flat space, $\TEMTE$ contributes to the $(p^2)^0$ structure.
We can therefore concentrate on $\TEMTO$.\footnote{In the case where the fermions are massive $\TEMTO \to  \TEMTO + N_f m_f (1 + \ga_m) \bar q q$.}
Adapting the notation $[O_\theta ] = \TEMTO$ in analogy with \eqref{eq:Og}, Eq.~\eqref{eq:TEMT} implies 
a relation between the two Laurent series
\begin{equation}
\label{eq:Lsg}
\lDR{\theta}{\RR} =    \frac{\hat{\be}^2}{4} \lDR{g}{\RR}  + \fin     \;.
\end{equation}
An expression for  $\lDR{\theta}{\MS} $ is obtained from \eqref{eq:Lgeps} by multiplying by $\hat{\be}^2/4$, partial integration 
and subtracting the finite constant in \eqref{eq:Lsg}
\begin{equation}
\label{eq:Lseps}
\lDR{\theta}{\MS}   =   \frac{1}{4}  \int_0^{\als}    \partial_u \left( \frac{\be}{u}   \right)u \left(\left( \frac{\mu(\als)}{\mu(u)} \right)^{2 \eps}- \frac{u}{\als} \right)  \rnDI{g}{g}{1}(u) du  \;.
\end{equation}
The limiting expression $\lDfR{\theta}{\MS}$ is manifestly finite and well-behaved in the limit 
$\be \to 0$ and $\als \to 0$. For instance,  the LL-expression is given by  
 $\lDfR{\theta}{\MS}|_{\textrm{LL}} = (\rnDI{g}{g}{1,0}/4 ) \cdot \be_0  \als$. 
 In passing we note that the criteria \eqref{eq:convAF} for the $\vev{\TEMTO\TEMTO}$-correlator is obeyed with $\ga_\theta =0$ and $n_{\theta\theta} = 2$. 
 Finiteness of 
 the trace of the $\vev{ \TEMTO\TEMTO}$-correlator and the $\vev{ G^2 G^2}$-correlator in AF gauge theories 
 has been noted elsewhere \cite{Zee81,Simon,RS09,LPR12} and \cite{Boch1}.
  
 We conclude this section by stating that both $ \cwiD{g} (p^2) $ and $ \cwiD{\theta} (p^2) $ 
 are finite for 
 $\eps \to 0$ and that  $ \cwiD{g} (p^2)$, being proportional to $1/\be$ \eqref{eq:Lgeps}, cannot 
 be expanded in $\als$. An AF theory is therefore different from a 
 conformal  field theory (CFT), of which a free theory is a special case, in 
 that the  correlator of  marginal operators is finite in the former but not the latter case. 
 In a CFT $\Gamma_{XX}(x^2) \sim 1/x^8$, with $\Delta_{O_X}=4$, which diverges upon Fourier transformation whereas $\Gamma_{gg}(x^2) \sim 1/x^8 f(\ln (\mu^2 x^2))$ converges in the AF case.

 \subsection{Consequences of Finiteness of $ \cwiD{g} (p^2) $ and $ \cwiD{\theta} (p^2) $}
 
 
There are three  points connected to  the finiteness of  $\lDR{g}{\RR}$ and $\lDR{\theta}{\RR}$ which we would like to discuss. 
First, since the bare Wilson coefficients $ \cwiD{g}(p^2)$ and $ \cwiD{\theta}(p^2)$ are both  finite,   
they satisfy  a dispersion relation which does not require subtractions (i.e. no regularisation). 
Note that if  regularisation was necessary then  the $\eps \to 0$ limit would not exist contrary to our findings.
 An explicit dispersion representation is given at LL in section \ref{sec:disp}. 
Second, since $ \cwiD{\theta}(p^2)$ is finite and scale independent (since bare) it may be 
related to a physical observable, which is indeed the case cf.  section \ref{sec:physical}.
A third aspect is the $R^2$-trace anomaly associated with the $\vev{\TEMTO \TEMTO} $-correlator.
Since anomalies \emph{can be} interpreted as originating from UV-divergences  one might wonder whether 
UV-finiteness means that the $R^2$-anomaly (related to $\vev{\TEMTO \TEMTO} $ cf. section  \ref{app:R2-anomaly})   is  an artefact of perturbation theory only.
 The answer to this question is no, at least in the $\MS$-scheme  since it is the 
$\ln \mu$-term  which is the  true signal   of the anomaly.  Finiteness though means that one can 
choose a scheme  \cite{PZprep} where the $R^2$-anomaly is absent or absorbed into the renormalisation 
of the dynamical operators.
 

\subsubsection{Explicit unsubtracted Dispersion Representation for leading Logarithms }
 \label{sec:disp}
 
We introduce  $P^2 \equiv -p^2$ where $P^2$ might be thought of as a Minkowski 
momentum allowing us to write the dispersion relation in the usual way.   
 The starting point is the LL expression \eqref{eq:LgLL}. 
The associated logarithms are $1/ \eps^{n}\leftrightarrow     - \ln^n(1/\mu^{2}) $ (which is derived in appendix 
\ref{app:LL} from the bare correlation function) 
and by dimensional analysis this 
implies $1/ \eps^{n}\leftrightarrow     -  \ln^n(-P^2/\mu^2) $.  
At LL the expression can be written as follows
\begin{eqnarray}
\label{eq:CWLLg}
\cwiD{\gsym}(p^2)\big|_{\textrm{LL}}   
&=& \cwiDR{\gsym}{\MS} (p^2)|_{\textrm{LL}}  +  \left( \lDR{g}{\MS}|_{\textrm{LL}} \right)_{\eps \to 0}  \nonumber  \\[0.1cm]
&=&  \left(\lDR{g}{\MS}|_{\textrm{LL}} \right)_{\epsilon^{-n} \to     -  \ln^n\left(-\frac{P^2}{\mu^2} \right)}
+ \left( \lDR{g}{\MS}|_{\textrm{LL}} \right)_{\eps \to 0} \nonumber  \\[0.1cm]
&=&    - \frac{\rnDI{g}{g}{1,0} \ln(-P^2/\mu^2)  }{1+ \als \be_0 \ln(-P^2/\mu^2) } + \frac{\rnDI{g}{g}{1,0}}{\als \be_0}   =  \frac{\rnDI{g}{g}{1,0}}{\als \be_0} x(P^2)  \;,
\end{eqnarray}
with 
\begin{equation}
\label{eq:x}
 x(P^2) =   \frac{1  }{1+ \als \be_0 \ln(-P^2/\mu^{2}) } \;,
\end{equation}
and it should be kept in mind that the $\eps^{-n}$ replacement rule is to be applied to the renormalised part only.  
We refer the reader to  \cite{Boch2} for an next-to-leading logarithmic  expression but the reason we content 
ourselves with LL is that it is sufficient for the asymptotic behaviour. 
Since $x(P^2)$ is finite for $P^2 \to \infty$, it obeys  an unsubtracted dispersion relation of the form
\begin{equation}
\label{eq:xdisp}
x(P^2) = \frac{1}{2 \pi i} \int_\Gamma \frac{ x(s)}{s-P^2} \;,
\end{equation}
where $\Gamma$ is such that no singularities are crossed. 
The singularities of $x(P^2)$ are a branch cut at  
 $P^2 \geq  0$ and a pole in the euclidean domain at $P^2 = P_0^2 \equiv - \mu^2
  \exp(-1/(\be_0 \als))  $ on which we comment in the next section.
It is convenient to  split the dispersion representation into the pole part 
\begin{equation}
\label{eq:xdisper}
x(P^2) = \frac{-1}{ 1 -      P^2/P_0^2}  + \hat{x}(P^2)\;, \end{equation}
and the integration over the cut
\begin{equation}
\label{eq:xh}
 \hat{x}(P^2) =  \frac{1}{\pi} \int_0^\infty ds \frac{{\textrm{ Im}}[x(s)]}{s-P^2-i0}   = 
  \int_0^\infty  \frac{ds}{s-P^2-i0} \frac{1}{(1+ a \beta_0 \ln (s/\mu^2))^2 + (a \beta_0 \pi)^2} \;.
\end{equation}
Above it was used that $x(s) \to 0$ for $s \to \infty$ as otherwise the arc at infinity would contribute to the dispersion integral. This is the formal solution and it is easily seen that for finite $P^2$ the integrand behaves 
 $\int_0^\infty  ds/( s \ln(s/\mu^2)^2) < 0$ which is finite.  The integral \eqref{eq:xh} is explicitly evaluated in 
 appendix \ref{app:check} to reproduce the expression in \eqref{eq:xdisper}. The dispersion relation for 
 the TEMT-part is simply given by  
 $ \cwiD{\theta} (p^2)\big|_{\textrm{LL}} = \be_0^2/4 \alsp{2}  \cwiD{g} (p^2) \big|_{\textrm{LL}}$.
  
\subsubsection{Finiteness  of $ \cwiD{\theta} (p^2) $  implies observability}
\label{sec:physical}

Generally physical quantities are RG-scale 
independent and finite. 
Bare correlation functions with renormalised composite operators, such as $\Gamma_{AB}(p^2)$ \eqref{eq:GAB}, are RG-scale independent but in the case where they 
are not finite,  do not  qualify as physical observables.  
Since $\cwiD{\theta}(p^2)$  is finite the situation changes and the bare function  is observable. For example  $\Delta \barb$,  the difference of the flow 
of $\Box R$-term of the Weyl anomaly,  is related  by  $ \Delta \barb = \frac{1}{8} \cwiD{\theta}(0)$ \cite{PZprep}.

Below we illustrate the scale independence of $\cwiD{\theta}(p^2)$ (and the analogous
case of the bare $m^2 \vev{\bar q q \bar qq}$-correlator is discussed in appendix \ref{app:mass}).
In a one-scale theory  with one external momentum any quantity reads  
$\varphi(p^2/\mu^2,\als(\mu/\mu_0))$ where 
$\mu_0$ is a reference scale, e.g. $\Lambda_{\textrm{QCD}}$, which we suppress further below. In the case 
where $\varphi$ is a physical quantity, and therefore independent of the renormalisation scale $\mu$, 
the functional dependence simplifies to 
\begin{equation}
\frac{d}{d \ln \mu} \varphi(p^2/\mu^2,\als(\mu^2)) =0 \quad \Leftrightarrow \quad \varphi = \tilde{\varphi}( \als(p^2)) \;.
\end{equation}
This is indeed the case for $ \cwiD{\theta} (p^2) $ at LL. Starting with 
\eqref{eq:CWLLg} one gets
\begin{equation}
\label{eq:CWLL}
\cwiD{\theta}(p^2) |_{\textrm{LL}} =  \frac{\rnDI{g}{g}{1,0}}{4}   \frac{ \als(\mu^2) \be_0  }{1+ \als(\mu^2)  \be_0 \ln(p^2/\mu^{2}) }  = 
\frac{\rnDI{g}{g}{1,0}}{4} \be_0 \als(p^2) + {\cal O}(\be_1) \;,
\end{equation}
a function which depends on  $\als(p^2)$  only. 
Note that if we were guided by fixed order perturbation theory then 
we would resort to the renormalised  $\cwiDR{\theta}{\RR} (p^2) $ 
which is scale dependent, $\frac{d}{d \ln \mu} \cwiDR{\theta}{\RR} (p^2) =  
- \lim_{\eps \to 0}  ( \frac{d}{d \ln \mu} - 2 \eps ) \lDR{\theta}{\RR} =   \zamDR{\theta}{\RR}$ with the last equality following from   \eqref{eq:LQu}.  This is why it is sometimes 
stated that only $p^2 \frac{d}{d p^2}\cwiDR{\theta}{\RR} (p^2) = 
p^2 \frac{d}{d p^2} \cwiD{\theta}(p^2)$ (e.g.  \cite{Zoller:2012qv}) is physical whereas we advocate that the bare term $\cwiD{\theta}(p^2)$ is physical and should be stated. 
An example being  the previously mentioned $\Box R$-flow: 
$\Delta \barb = \frac{1}{8} \cwiD{\theta}(0)$.
The scheme dependent splitting of the bare function into a counterterm and a renormalised part  defines a flow for $\barb$ connecting the UV and IR values \cite{PZprep}.

The pole discussed in the
previous section is  the Landau pole of the gauge coupling.  It has no direct physical meaning and contradicts the analytic structure of the spectral representation.
 It is precisely this pole that is removed  in the approach of analytic perturbation theory by enforcing a physical singularity structure 
on the amplitude \cite{aPT}.  
In the fully non-perturbative version this pole disappears. 
The correlation function satisfies an unsubtracted dispersion ($P^2 = - p^2$)  
\begin{equation}
\label{eq:xhS}
 \cwiD{\theta}(P^2)  =  \frac{1}{\pi} \int_0^\infty ds \frac{{\textrm{Im}}[\cwiD{\theta}(s)]}{s-P^2-i0} + \omega_0     \;,
\end{equation}
consistent with the K\"all\'en-Lehmann representation.
Above $\omega_0$ is an arbitrary, scale independent, finite constant which 
can be added by changing the theory by a local term in the UV.  
The addition of this term is more than a choice of scheme, it corresponds to changing the
theory by a local term. We have therefore silently assumed $\omega_0 =0$ which is automatic in the conventional setup.
  This constant, being arbitrary, should not impact on any physical predictions.   
 In the above mentioned formula of the $\Box R$-flow anomaly this is ensured 
 by the implicit boundary condition  $\cwiD{\theta}(\infty)  = 0$. If this boundary condition is generic $\cwiD{\theta}(\infty)  = \omega_0$, as in \eqref{eq:xhS}, then the formula simply changes 
 to $\Delta \barb = \frac{1}{8}( \cwiD{\theta}(0) - \omega_0)$ \cite{PZprep}.

\subsection{OPE-extension with Condensates}
\label{sec:OPE}

This section may be considered as a minor digression and the reader may or may not want 
to directly proceed to section \ref{sec:extension}. The discussion below has some overlap 
with reference \cite{S91} but goes beyond it in the emphasis on finiteness. 
So far we have treated the correlation function \eqref{eq:GAB} within the framework of  perturbation theory. 
Eq.\eqref{eq:GAB} is a good approximation for large $p^2$ and 
 presasymptotic effects for $p^2 \gg \Lambda_{\textrm{QCD}}^2$  
 can be parameterised in terms of vacuum condensates $\vev{[O_C]} \sim {\cal O}(\Lambda_{\textrm{QCD}}^4)$ with the framework of the OPE \cite{Wilson:1969zs}.
 The vacuum condensates appear as power suppressed wrt  perturbation theory 
 (i.e. the $\vev{\ONE} = 1$ term)  
\begin{equation}
\Gamma_{AB}(p^2)    =
 \cwi{A}{B} (p^2)  p^4 
  \vev{\ONE}    +   \sum_{C}  \CWI{A}{B}{C}(p^2) \vev{[O_C]} \;.
\end{equation} 
Anticipating section \ref{sec:multiple} we include several operators which correspond 
to the several coupling case.
In the formula above 
we have assumed all perturbations $[O_C]$ to be of dimension four, i.e. marginal.  
The OPE has been shown to hold in perturbation theory \cite{OPEZ,Wilson:1972ee}
and has enjoyed phenomenological success 
in the non-perturbative regime where it is expected to hold \cite{Novikov:1984rf}.

It is our aim to investigate whether or not the Wilson coefficient $\CWID{g}{g} (p^2)$, in analogy to $\cwiD{g} (p^2)$,
\begin{equation}
\Gamma_{gg}(p^2)   =  \cwiD{g} (p^2) p^4 +  \CWID{g}{g} (p^2)  \vev{[O_g]} \;,
\end{equation}
is finite or not. To do so the  QAP  proves useful.  
The key idea of the QAP is that differentiation of the finite partition function with respect to finite 
renormalised parameters leads directly to finite well-defined quantities.  The QAP might be regarded 
as a scheme for the renormalisation of composite operators.
For convenience we employ the local QAP (e.g. \cite{S91,S16}) where  the couplings $\gc{A}$  are promoted to local functions $ \gc{A}(x)$
which then become sources for  the corresponding local  operators
\begin{equation}
\label{eq:conjugate}
\vev{[O_A(y)]} =  (- \Dloc{A}{y})  \ln \Zpart   \;, \quad \Dloc{A}{x}  \equiv  \frac{\de}{\de  \gc{A}(x)}  \;,
\end{equation}
where the corresponding Lagrangian in bare quantities reads ${\cal L} = \gcb{A} O_A + \dots$ .
The principle also applies 
to  higher point functions such as
\begin{eqnarray}
\label{eq:main}
\Gamma_{AB}^{\RR}(x-y) &\; =\;& (-\Dloc{B}{y}) (-\Dloc{A}{x})  \ln \Zpart  =
(-\Dloc{B}{y})  \vev{ [O_A(x)]}  \nonumber \\[0.2cm] 
 &\; =\;&  \vev{ [O_A(x)] [O_B(y)] }_c  -
\vev{ \left(\Dloc{B}{y}[O_A(x)] \right)}  = \fin \;.
\end{eqnarray}
Since the right hand side is finite this means that the local divergences of the 
unrenormalised $2$-point function
\begin{equation}
\label{eq:start}
\CWI{A}{B}{C}(p^2) -  \LnDR{A}{B}{C}{\RR} = \fin \;,
\end{equation}
 ought to cancel against  corresponding divergencies in $\LnDR{A}{B}{C}{\RR}$ coming from the variation of the operator renormalisation constants
\begin{equation}
\label{eq:contact}
 \vev{ \left(\Dloc{B}{y}[O_A(x)] \right)} 
 =   \sum_{C} \LnDR{A}{B}{C}{\RR}  \vev{[O_C]}  \delta(x-y) +
\mu^{d-4}  \lnDR{A}{B}{\RR}   \vev{\ONE} \Box^2  \delta(x-y)+\fin \;.
\end{equation}
The quantities $\LnDR{A}{B}{C}{\RR}$ are given in terms of the RG mixing matrix $\ZopI{A}{I}$
\begin{equation}
\label{eq:RenContact}
\LnDR{A}{B}{C}{\RR} = ( \Dglob{B} \ZopI{A}{I} )  \ZopII{I}{C}  \;, \quad \Dglob{Q} \equiv  \frac{\partial}{\partial g^Q}  \;,
\end{equation}
where the scheme dependence comes from the scheme-dependent $\ZopI{A}{I}|_{\RR} = \ZopI{A}{I}$ which we suppress throughout in order to lighten the notation. 
For our example with a Yang-Mills Lagrangian  ${\cal L} = 
1/4 O_g$ with $O_g$ defined in \eqref{eq:Og} the renormalised composite operator 
follows from 
$ \vev{[ O_g(x) ] } = 
(- \frac{4}{g^2}\delta_{1/g^2(x)} ) \ln \Zpart  = ( 2 \de_{\ln g(x)}) \ln \Zpart $.  
The renormalisation of $[O_g]$ is given by \cite{KlubergStern:1974rs,Spiri84}
\begin{equation}
\label{eq:OgR}
[O_g] =  \ZopI{\gsym}{\gsym}  O_g +  \dots  \;, \quad \ZopI{\gsym}{\gsym} =
 (1+ \frac{\partial \ln Z_g}{\partial \ln g}) = \frac{(d-4)}{2\hat{\beta}} \;,
\end{equation}
with the dots standing for \eom\,and gauge dependent terms which 
vanish on physical states and are therefore immaterial for the current discussion. 
From eqs.~(\ref{eq:contact},\ref{eq:OgR}) one gets
\begin{eqnarray}
\label{eq:gggContact}
 \LDR{g}{g}{\Ra} &\!=\!& 2 \partial_{\ln g}  \ln  \ZopI{\gsym}{\gsym}  
  \nonumber \\[0.2cm] 
 &\!=\!&        
  2 \left(1+ \frac{\partial \ln Z_g}{\partial \ln g}\right)^{-1}   \frac{\partial^2 \ln Z_g}{\partial (\ln g)^2}         
    \nonumber \\[0.2cm] 
 &\!=\!&     - \frac{2 \partial_{\ln g} \hat{\be}}{\hat{\be}}  
 \stackrel{\eps/\be \to 0}{\to}  \fin 
  \;.
\end{eqnarray}
The   term $\LDR{g}{g}{\Ra}$ differs from  $\LDR{g}{g}{\MS}$ in finite terms since the latter 
is a  power series in $1/\eps$ by definition. From \eqref{eq:start} it then follows that $\CWID{g}{g} (p^2)$ is finite in the limit  $\eps \to 0$ 
but divergent at each order in perturbation theory.\footnote{To compare with the literature we need the CTs of $G^2$ as opposed to $[O_g]$. These can be obtained by accounting for the factor $1/g_0^2$ in the definition of $O_g$ (see \eqref{eq:Og}), which leads to the following modification of \eqref{eq:gggContact}
$
  2 \partial_{\ln g}  \ln  \ZopI{\gsym}{\gsym} /g_0^2 =  \LDR{g}{g}{\Ra}    -  2(d-4)/\hat{\be}
$.
 Expanding this expression in powers of $\als$ results in
$ -4 \left( \alsp{2}\left[    \frac{\beta_1}{\epsilon} \right]  + \alsp{3}  
 \left[    - \frac{ \beta_0 \beta_1}{\epsilon^2}  + \frac{2 \beta_2}{\epsilon}    \right]  +  
 \alsp{4}  \left[\frac{\be_0^2 \be_1}{\epsilon^3}   - \frac{\be_1(\be_1+ 2 \be_0)}{\epsilon^2}  
 + \frac{3 \be_3}{\epsilon}  \right] \right)  $ 
 up to corrections of the order of  $  {\cal O}( \alsp{5} ,\epsilon^0 ) $.
 This is identical to  $\frac{1}{4}Z_{11}^L/Z_{11}$ in \cite{Zoller,Zoller:2016iam} [Eq.4.7] 
up to finite terms. However, the observation of finiteness and its possible implications were not made 
in \cite{Zoller,Zoller:2016iam}.}  
It is noted though that  
$\LDR{\theta}{\theta}{\RR} (p^2) = \hat{\be }/2 \LDR{g}{g}{\RR} + \fin  =  - \partial_{\ln g} \hat{\be} + \fin$  is finite, but even finite in each order 
in perturbation theory provided every quantity, e.g. beta-function, is treated consistently in $d$-dimensions. 
The same applies 
therefore to the corresponding bare Wilson coefficient $\CWID{\theta}{\theta} (p^2) = \hat{\be }/2 \CWID{g}{g} (p^2)$.
Hence one can  write down convergent dispersion relations for both 
$ \CWID{g}{g} (p^2)$ and $\CWID{\theta}{\theta} (p^2) $ as done in section \ref{sec:disp}. 
The coefficient $\CWID{\theta}{\theta} (p^2) $ is scale independent. 

It is again instructive to write down the LL expression 
\begin{alignat}{2}
& \CWID{g}{g}(p^2)|_{\textrm{LL}} &\;=\;&  \frac{ -1  }{1+ \als(\mu^2) \be_0 \ln(p^2/\mu^{2}) } \;, \nonumber \\[0.1cm]
& \CWID{\theta}{\theta}(p^2)|_{\textrm{LL}} &\;=\;&  \frac{ \be_0 \als(\mu^2)   }{2(1+ \als(\mu^2)   \be_0 \ln(p^2/\mu^{2}) )} 
= \frac{\be_0}{2} \als(p^2) + {\cal O}(\be_1) \;,
\end{alignat}
which happens to be proportional to $\cwiD{g}(p^2)|_{\textrm{LL}}$ and 
$\cwiD{\theta}(p^2)|_{\textrm{LL}}$   \eqref{eq:CWLL} respectively.

\subsection{Extending the $R^2$-anomaly to ${\cal O}(\alsp{5})$}
 \label{app:R2-anomaly}
 
 
 \subsubsection{$R^2$-anomaly from the $\vev{\TEMTO\TEMTO}$-correlator  and the Quantum Action Principle}
 
 It has been known for a long time that  trace anomalies in curved space time 
 (e.g. \cite{H81})
 are related to correlation functions of the TEMT in flat space. The link is provided 
 by the local QAP. 
 
In four dimensions the  VEV of the TEMT in curved space reads (e.g. \cite{birrell1982quantum})
\begin{equation}
\label{eq:VEVTEMT}
\vev{\TEMT{\rho}} = -( \ba^\IR E_4 + \bc^\IR W^2  + \bb^\IR H^2)  + 4 \barb^\IR \Box  H  + d \Lambda^\IR \;, \quad H \equiv \frac{1}{(d-1)} R   \;,
\end{equation}
with $E_4$, $W^2$, $R$ and $\Lambda$ being the Euler, the Weyl squared   the Ricci scalar 
and the cosmological constant term and $  \ba(\mu) \stackrel{\mu \to 0}{\to} \ba^\IR$ etc.\footnote{The latter is defined by $ \vev{T_{\al \be}} = 
g_{\al\be} \Lambda^\IR $ and  
may or may not be canceled by adding a suitable counterterm to  the UV action.} 
The Euler term is topological and analogous to the $G \tilde G$-term 
of the chiral anomaly known as a type A anomaly \cite{DS93}. 
The $W^2$ and $H^2$ arise from the introduction of a scale in the process of regularisation 
and $\Box H$ is the variation of a local term. These are known as type B anomalies \cite{DS93}. 

We write  the gravitational counterterm as follows  
 \begin{equation}
 \label{eq:Lgrav}
 {\cal L}_{\textrm{gravity}} = -( a_0 E_4 + c_0 W^2 +   b_0  H^2)   \;,
 \end{equation}
with $b_0 = \mu^{d-4} (b^\RR + L_b^\RR)$ and notation that largely  follows  
Shore's review  \cite{S16}. 
 A double variation of the Weyl-factor $s(x)$  ($g_{\mu\nu} \to e^{ - 2 s(x)} g_{\mu\nu}$) is finite 
 since both the partition function and the metric are finite. 
When the latter is Fourier transformed and projected on the $p^4$-structure 
one obtains
\begin{eqnarray} 
\label{eq:2ptEM}
 \int d^d xe^{i p \cdot x} \left(  \left(-{\delta_{s(x)}} \right) \left(-{\delta_{s(0)}} \right) \ln \Zpart \right)  |_{p^4} 
 &=& 
  \int d^d x e^{i p \cdot x}  \vev{ \TEMT{\rho} (x) \TEMT{\rho}(0)}|_{p^4}+ 8  \, b_0   
  \nonumber \\[0.1cm]  
 &=&    \int d^d x e^{i p \cdot x}  \vev{ \TEMTO(x)  \TEMTO(0)}|_{p^4}+ 8  \, b_0  
  \nonumber \\[0.1cm]  
  &=& \cwiD{\theta}(p^2) p^{-2\eps}  +   8  \, b_0    =  \fin \;,
 \end{eqnarray} 
 where in passing to the third line we used the fact that eom terms contribute to the $(p^2)^0$-structure only and assumed that  
neither virial currents or non-improvable scalars are present.  This is correct 
for QCD-like theories in the conformal window used in the next section.  
In the parameterisation of \eqref{eq:link} this implies the non-trivial, known \cite{F83}, relation 
\begin{equation}
\label{eq:link}
L_b^\Ra  =- \frac{1}{8}\lDR{\theta}{\Rb}  + \fin \;,
\end{equation}
which translates into $L_b^\MS  =- \frac{1}{8}\lDR{\theta}{\MS}$ for the $\MS$-scheme.
We end this section with three slightly disjoint points 
\begin{itemize}
\item One application of  the finiteness of $L_b^\RR$, i.e. $p^4$-structure of the 
$\vev{\TEMTO \TEMTO}$-correlator, is that  one can choose a scheme where
$\bb^\RR$ vanishes \cite{PZprep}. The contribution is absorbed into the operators appearing 
in the trace anomalies e.g. $G^2$ in the case of QCD-like theories.
Below $\bb$  is given in the $\MS$-scheme for which is non-zero outside the FPs.
\item Remark on the sign conventions and the specifics of the gravity counterterms. First  $a_0$, $b_0$ and $c_0$ are taken 
to be independent of the scale in accordance with references \cite{BC80,H81,F83} but differing from the classic work \cite{JO90} where $b_0$ is reduced to $\mu^{d-4}L_b^\MS$. We refer the reader 
to appendix B of our paper \cite{PZprep} for further comments.
Our approach determines $b^\RR$ in $b_0 = \mu^{d-4}(b^\RR+L_b^\RR)$ up to 
a scale independent constant  e.g. \cite{H81} which  incidentally is
the $\omega_0$ in \eqref{eq:xhS}.
The sign convention of $b_0$ is such that $\be_a$ decreases along the flow 
leading to the counterterm with the opposite sign in \eqref{eq:2ptEM} as compared
to \eqref{eq:xspace}, which explains the  sign difference in \eqref{eq:link}.
\item Eq. \eqref{eq:link}  allows us to elucidate  
 the quantity $\zamDR{Q}{\RR}$ \eqref{eq:RGE1} in the context where $Q=\theta$.  
 Since the TEMT is physicsal, $\ga_\theta = 0$, and from \eqref{eq:LQu}  it follows that 
 (  $\be_b^\RR    \equiv -  (  \frac{d}{d \ln \mu}  - 2\eps )   L_b^\RR $)
  \begin{equation}
 \label{eq:link2}
   \be_b^\MS  =   - \frac{1}{8} \zamR{\theta}{\theta}{\MS} \;,
 \end{equation}
$\zamR{\theta}{\theta}{\MS}$  and the $ \bb^\MS R^2$-anomaly  \eqref{eq:VEVTEMT} are related. 
It seems worthwhile to mention that 
the link between $\zamR{A}{B}{\RR}$ and the TEMT  generalises for 
 local source-couplings other than $s(x)$.
Instead of geometric terms like $R^2$, the $\zamR{A}{B}{\RR}$ appear in front of covariant expression 
in the source couplings $g_{A,B}(x)$ cf. [Eqs 2.5,2.8] in \cite{JO90}. The concept also 
generalises to higher point functions both at the level of gravity terms (e.g. $E_4$ is related
 to $3$-point functions) as well as covariant coupling terms.
\end{itemize}

\subsubsection{Application to QCD-like Theories}

The generally valid relations (\ref{eq:link},\ref{eq:link2})  are applied to QCD-like theories 
in this section.
From \eqref{eq:link} and $\lDR{\theta}{\MS}$ in \eqref{eq:Lseps} it is then observed  that  ($\eps \to 0$   implied)  
\begin{eqnarray}
\label{eq:Lbmu}
L_b^\MS(\als(\mu))  =  - \frac{1}{32}  \int_0^{\als}    \partial_u \left( \frac{\be}{u}   \right)u \left(1- \frac{u}{\als} \right)  \rnDI{g}{g}{1} (u) du \;,
\end{eqnarray} 
 which allows us to write an explicit formula for the $\be_b$-anomaly term
 \begin{equation}
\label{eq:beb-g1}
\be_b^\MS = -  \left(  \frac{d}{d \ln \mu}  - 2\eps \right)   L_b^\MS =  
\frac{1}{16}  \frac{\beta(\als)}{\als} \int^{\als}_0 \partial_u \left( \frac{\beta(u)}{u} \right)   u^2 \rnDI{g}{g}{1} (u) du  \;.
\end{equation}
 Hence from the $\rnDI{g}{g}{1}$ counterterm of the TEMT correlation function one can deduce 
 the $R^2$-anomaly term.
The $\rnDI{g}{g}{1}$-term can be found in the recent computation   \cite{Zoller:2016iam} 
up to NNLO. We extract
\begin{eqnarray}
\label{eq:gex}
\rnDI{g}{g}{1,0}  \!\!\! \!  &=& \!\!\! \! \frac{n_g}{4 \pi^2}  \;,   \\[0.1cm]  
 \rnDI{g}{g}{1,1} \!\!\! \!  &=&  \!\!\! \! \rnDI{g}{g}{1,0}      \left( \frac{17}{2} C_A - \frac{10}{3} N_F T_F \right)     \;, \nonumber   \\[0.1cm] 
 \rnDI{g}{g}{1,2} \!\!\! \! &=& \!\!\! \! 4 \rnDI{g}{g}{1,0}  \left(C_A^2(\frac{11}{6} \zeta_3 + \frac{22351}{1296})-C_A N_F T_F(\frac{14}{3} \zeta_3+\frac{799}{81} )+ n_g N_F T_F (  \zeta_3-\frac{107}{18})+\frac{49}{81} T_F^2 N_F^2  \right)   \;,
 \nonumber
\end{eqnarray}
with $\zeta_3$ being the Riemann zeta function at the value $3$ and $n_g = C_A C_F /T_F |_{SU(N_c)} = N_c^2-1$ being the number of gluons (dimension of the adjoint representation).
The $ {\cal O}(\alsp{3})$ contribution agrees with \cite{F83} [Eq.7.7] at the level of $\be_0$ and $\be_1$ 
which straightforwardly extends  to QCD-like theories. 
 
 From \eqref{eq:beb-g1} one then obtains $\bb$ up to ${\cal O}(\alsp{5})$. We give the result in terms of the first pole 
 in $L_b^\MS = \sum_{n \geq 1} b_n(\als) \eps^{-n} $ by the $\MS$-type relation (with an  extra factor of  $2$ in the first equality wrt \cite{H81,F83} 
from the $d = 4- 2 \epsilon$ versus $d= 4- \eps$ convention)
 \begin{equation}
\label{eq:c-anomaly}
\bb^\MS = 2  \partial_\als  (\als b_1) = 8\Bone{3} \alsp{3}  + 10 \Bone{4} \alsp{4} + 12 \Bone{5} \alsp{5} +   {\cal O}( \alsp{6}) \;,
\end{equation}
 where ($b_n = b_{n,0} + b_{n,1}  \als  + {\cal O}(\alsp{2})$)
\begin{eqnarray}
\label{eq:C1i}
 \Bone{3} &=&  \frac{1}{24\cdot 16} \beta_0 \beta_1 \rnDI{g}{g}{1,0}    \;, \quad 
  \Bone{4} =   \frac{1}{120\cdot 16} (   (4\beta_{1}^2   
+  6 \beta_{0} \beta_{2}) \rnDI{g}{g}{1,0}  + 3 \beta_0 \beta_1 \rnDI{g}{g}{1,1} )  \;, \nonumber 
\\[0.2cm]
\Bone{5} &=&   \frac{1}{720 \cdot 16 } (50 \be_0 \be_2 \rnDI{g}{g}{1,0}   + (24 \be_0 \be_2  + 15   \be_1^2) \rnDI{g}{g}{1,1}   + 12    \be_0 \be_1 \rnDI{g}{g}{1,2}   )   \;,
\end{eqnarray}
follows from \eqref{eq:beb-g1}.
Comparing with  \cite{H81,F83,JO90}  
we find agreement with \cite{H81,F83} to the computed order of $ {\cal O}(\alsp{3})$ and with \cite{JO90} 
to order $ {\cal O}(\alsp{4})$. 
The fact that $\bb^\MS$ is proportional to  the $\be$-function is consistent 
with $\bb$ being zero in CFTs \cite{D77,BCR83}. In fact a stronger statement 
can be made since 
formula \eqref{eq:beb-g1}
shows that  $\bb^\MS$ vanishes for one-loop $\be$-functions consistent with 
all coefficients above involving  a $\be_n$ with $n \geq 1$.

At least let us make a general observation which makes us of the finiteness of $\bar{L}^\MS_b = 
\lim_{\eps \to 0} L^\MS_b$. 
Firstly, we observe that by one may take the $\eps \to 0$ limit in \eqref{eq:beb-g1} directly and 
replace $\frac{d}{d \ln \mu } \to  \be \frac{\partial}{\partial \ln g}$ leads to $\be_b^\MS = 
- \be \frac{\partial}{\partial \ln g}\bar{L}^\MS_b$. This result generalises to multiple couplings as follows
\begin{equation}
\label{eq:B}
\be_b^\MS = 
- \be^A  \partial_A \bar{L}^\MS_b \;,
\end{equation}
where $\bar{L}^\MS_b$ is well-defined in the limit of vanishing $\be$-function as will be shown in 
section \ref{sec:multiple}. This result is accordance with $\be_b^{\text{CFT} }= 0$ \cite{D77,BCR83}.

\section{Extensions of the one-coupling $2$-point Function Case}
\label{sec:extension}

\subsection{Multiple Couplings and Finiteness of TEMT-correlators}
\label{sec:multiple}


In this section we proceed to show the finiteness of the $\VEV{\TEMTO \TEMTO}$-correlator for a  general field theory  with an UV-FP. Consider a RG-flow generated by a deformation 
$\delta {\cal L} = \sum_A g^A_0 O_A$. The induced trace anomaly 
reads\footnote{Three possible structures are neglected. EOM-terms can be omitted 
for the same reasons as before. It is assumed that no virial currents  
$  \TEMTO = \partial \cdot V + \dots$ are present implicit in the assumption that the UV-FP 
is conformal (no non-trivial unitary scale but not conformally invariant theories are known 
to date). Terms of the form $ \TEMTO = - \Box \phi^2 + \dots $ originating from non-conformally 
coupled scalars can be improved  \`a la Callan, Coleman and Jackiw \cite{CCJ70}. 
An exception is the chirally broken phase but since the term is relevant in the IR and not the UV we do not need to consider it for the purposes of this section.}
 \begin{equation}
\label{eq:MultTEMT}
\TEMTO = \hat{\be}^A [O_A]    \;,
\end{equation}
where here the $\be$-functions for the couplings $g^A$  are given  by
\begin{equation}
\hat{\be}^A= \frac{d}{d \ln \mu}  g^A   = \be^A - \eps g^{A} \xi^A  \;.
\end{equation}
The $\xi^A$ are an artefact of going from $4$ to $d$ dimensions (e.g. \cite{JO90} whose notation is adapted here).
Note that  in section \ref{sec:QCD-like}, unlike here, the logarithmic $\be$-function was used  
and that in QCD  $\xi^g = 1$.
The generalisation of \eqref{eq:QQfin} to the non-diagonal case is straightforward and given by
\begin{equation}
\label{eq:2ptMult}
 \cwi{A}{B} (p^2)  p^{-2 \eps}   =   
  \big( \cwiR{A}{B}{\RR} (p^2)   +  \lnDR{A}{B}{\RR}  \big)  \mu^{-2 \eps}   \;,
\end{equation}
where $\cwi{A}{B}$ is again finite in the sense that there are no poles 
upon expanding in $\eps$ as long as no expansion in the couplings $g^Q$ is attempted. 
 The multiple coupling generalisation of 
  \eqref{eq:RGE1} reads
\begin{equation}
\label{eq:RGEm}
(  {\cal L}_\be  - 2 \eps)  \lnDR{A}{B}{\RR} = -  \zamR{A}{B}{\RR} \;,
\end{equation}
where ${\cal L}_\be$ denotes the Lie derivative on a $2$ tensor in coupling space
\begin{equation}
{\cal L}_\be  \lnDR{A}{B}{\RR}  = 
\partial_A \hat{\be}^C  \lnDR{C}{B}{\RR}+\partial_B \hat{\be}^C  \lnDR{A}{C} {\RR}+ 
 \hat{\be}^C \partial_C \lnDR{A}{B}{\RR} \;,
\end{equation}
($\partial_A$  defined in \eqref{eq:contact}) since  $\hat{\ga}_{A}^{\phantom{A} B}$ is
\begin{eqnarray}
\label{eq:galog}
\hat{\ga}_{A}^{\phantom{A} B}  &=&  \partial_A \hat{\be}^B \nonumber \\[0.1cm]
&=& \partial_A \be^B  - \delta_{A}^{\phantom{A} B} \xi^A  \eps =  \gamma_{A}^{\phantom{A} B}  - \delta_{A}^{\phantom{A} B} \xi^A  \eps \;,
\end{eqnarray}
which  follows from  
$\frac{d }{d \ln \mu}\vev{\TEMTO} = 0$ in flat space.
The above equation is  the analogue of  $\ga_g = \hat{\ga}_g =  2   \plnas   \hat{\be}$ stated below \eqref{eq:Og}. The reason for $\ga_g = \hat{\ga}_g$ is that we used the logarithmic 
$\be$-function for QCD-like theories for which the ${\cal O}(\eps)$-term is coupling independent.
The quantity $\zamR{A}{B}{\MS}$ generalising  \eqref{eq:RGE1} is then given by
\begin{equation}
\label{eq:chiAB}
\zamR{A}{B}{\MS}  = 2\left(1+\frac{1}{2}(\xi^{A}+\xi^{B})+\frac{1}{2} \xi^{Q} g^Q \partial_Q \right) \rnDI{A}{B}{1} \;, \quad \lnDR{A}{B}{\MS} = \sum_{n \geq 1} \frac{\rnDI{A}{B}{n} }{\eps^n} \;,
\end{equation}
The RGE \eqref{eq:RGEm} can be solved  by the method of characteristics in terms of the anomalous dimension matrices $\ga_{A}^{\phantom{A} B}$
\begin{equation}
\label{eq:LQtMult}
\lnDR{A}{B}{\RR}(\mu) =  \int_{\ln \mu}^\infty I_{A}^{\phantom{A} C}(\mu,\mu')\zamR{C}{D}{\RR}(\mu') I_{B}^{\phantom{B} D}(\mu,\mu') \left( \frac{\mu}{\mu'} \right)^{2 \eps}  d \ln \mu'  \;, 
\end{equation}
where
\begin{equation}
\label{eq:IAB}
I_{A}^{\phantom{A} B}(\mu,\mu')  = ( \exp \left(  \int_{\ln \mu}^{\ln \mu'} \hat{\ga}(\mu'') d \ln \mu''   \right) )_{A}^{\phantom{A} B} \;.
\end{equation}
It can be shown that\footnote{This follows by writing $\hat{\be}^A(\mu) I(\mu,\mu')_A^{\phantom{A}B}= f^B(\mu')$ which satisfies the differential equation $\partial_{\ln \mu'} f^B= f^C\hat{\gamma}_{C}^{\phantom{C} B}(\mu')$ with initial condition $f^B(\mu)= \hat{\be}^B(\mu)$. It is easy to show using \eqref{eq:galog} that $f^B(\mu')=\hat{\be}^B(\mu')$ is the unique solution to to the initial value problem. }
\begin{equation}
\hat{\be}^A(\mu) I(\mu,\mu')_{A}^{\phantom{A} B}= \hat{\be}^B(\mu') \;.
\end{equation}
 As previously $\cwiD{\theta}(p^2)= \hat{\be}^A \hat{\be}^B \cwi{A} {B} (p^2)$ and the generalisation of \eqref{eq:Lsg} reads
\begin{equation}
\label{eq:MultiPoles}
\lDR{\theta}{\RR}(\mu) =  \int_{\ln \mu}^\infty \hat{\be}^A (\mu') \hat{\be}^B(\mu') \zamR{A}{B}{\RR}(\mu')  \left( \frac{\mu}{\mu'} \right)^{2 \eps}  d \ln \mu'  + \fin \;.
\end{equation}
For the asymptotic analysis it is, again, more convenient to use the variable $t = \ln \mu'$.
 We will now argue that the $\eps \to 0$ limit can be safely taken. 
Assuming $\zam{A}{B} = {\cal O} ( t^{-n_{AB}})$  with $n_{AB} \geq 0$   
 the integrand of \eqref{eq:MultiPoles} is controlled by the 
$\be$-functions for large  $t$ which tend to $0$ by the UV-FP assumption. 

Note that the criteria for finiteness are easily generalised.  
Finiteness of $\lDR{\theta}{\RR}$ and $\lnDR{A}{B}{\RR}$ is easily established. 
For the cases of AF and AS, of sections \ref{sec:AF} and \ref{sec:AS} respectively,  
 $\be^Q_{\textrm{AF}}= - \be^Q_0/ (4 \pi)^2 (g^Q)^{1+r_Q}+...$ and $r_Q>0$ and  $\be^Q_{\textrm{AS}} = |a_Q|((g^{Q})^\UV - g^Q) + ...$. 
 Expressed in the RG time variable $t$ this reads
 \begin{equation}
 \label{eq:CouplingBehaviour}
\be_{\textrm{AF}}^Q \sim \frac{1}{t^{(1+\frac{1}{r_Q})}} \;, \quad \hat{\be}_{\textrm{AS}}^Q \sim 
e^{- |a_Q| t} \;.
 \end{equation}
This means that the terms in the integrands in \eqref{eq:LQtMult} and  \eqref{eq:MultiPoles}  vanish at least as 
$t^{- (2+\frac{1}{r_A} + \frac{1}{r_B}   )}$ or are exponentially suppressed which guarantees convergence of the $t$-integral.
Hence the $\eps \to 0$ limit can be taken safely and $\lDR{\theta}{\RR}$ and $\lnDR{A}{B}{\RR}$  are finite which is the aimed result.  Note that  $\lnDR{A}{B}{\RR}$ is the analogue 
of $\lnDR{g}{g}{\RR}$ in the QCD-like case. If the operators are part of the dynamics i.e. 
present in the trace anomaly then the case has to be reconsidered.


\paragraph{Condensate corrections:}  As before we proceed to discuss the finiteness in the presence of vacuum condensates. The local coupling formalism of section \ref{sec:OPE} applies. 
  From \eqref{eq:start} and \eqref{eq:MultTEMT} it follows that
\begin{equation}
\label{eq:TTCcontact}
\LnDR{\theta}{\theta}{Q}{\RR}= \hat{\be}^A \hat{\be}^B \LnDR{A}{B}{Q}{\RR} + \fin \;.
\end{equation}
Along with \eqref{eq:RenContact} one deduces 
\begin{equation}
\hat{\be}^A \hat{\be}^B \LnDR{A}{B}{Q}{\RR} =\hat{\be}^A\hat{\be}^B( \Dglob{B} \ZopI{A}{P} )  \ZopII{P}{Q}=  -\hat{\be}^A \hat{\ga}_{A}^{\phantom{A} Q}=\fin \;,
\end{equation}
because of the boundedness of the anomalous dimension matrix $\hat{\ga}_{A}^{\phantom{A} Q}$. The scheme-dependence on the left hand side arises from $\ZopII{P}{Q} = \ZopII{P}{Q}|_{\RR}$ which we suppress. 
Finiteness of $\LnDR{\theta}{\theta}{Q}{\RR}$ follows  from \eqref{eq:TTCcontact}  and completes the task of this paragraph.

\subsection{Finiteness Criteria for $3$-point Functions}
\label{sec:3ptMom}

Another extension of interest are higher point functions. 
In general they consist of kinematic structures which are sensitive to the anomalous dimension 
of all operators in the correlation function  \emph{and} structures which are governed by lower dimensional point functions. 
The latter can be identified by setting one or more of the external momenta to zero. 
A comprehensive analysis in CFTs can be found in \cite{Skenderis} whereas we focus 
on theories with a non-trivial flow. 

We introduce 
the following notation 
\begin{eqnarray}
\label{eq:GABC}
\Gamma_{ABC}(p_A^2,p_B^2,p_C^2)  &=&   
 \int d^4 x  d^4 y  e^{i (p_A \cdot x + p_B \cdot y) } \vev{[O_A(x)][ O_B(y)][ O_C(0)]}_c   \nonumber  \\[0.1cm]
   &=&  \cwi{(A)}{BC} (p_Q^2)  p_A^4   +\cwi{A}{BC} (p_Q^2)  P_{BC} + \text{cyclic} \;,
 \end{eqnarray}
where cyclic permutation over $A$, $B$ and $C$ is implied, $p_A + p_B + p_C =0$, 
$Q = A,B,C$  
and $P_{BC} = p_A^4 - p_A^2(p_B^2+p_C^2)$ are kinematic structures which 
vanish when \emph{any} of the three external momenta $p_{A,B,C}$ is set to zero. 
Hence the $\cwi{(A)}{BC}$-coefficients  are the 2-point functions structures.
By applying 
\begin{equation}
\partial_{B} \cwiR{A}{C}{\RR} (p_Q^2)+\partial_{C} \cwiR{A}{B}{\RR} (p_Q^2)-
\partial_{A} \cwiR{B}{C}{\RR} (p_Q^2) = \fin \;,
\end{equation}
using the global version of \eqref{eq:contact} and noting  $\partial_{A}$ corresponds  to a 
zero momentum insertion of $[O_A]$,   the following equation for the Laurent series emerges
\begin{equation}
L_{(A)BC}^{\ONE,{\RR}}  =  L^{Q,{\RR}}_{BC}\lnDR{Q}{A}{\RR} - \frac{1}{2}(\partial_{B}\lnDR{A}{C}{\RR} +\partial_{C}\lnDR{A}{B}{\RR}-\partial_{A}\lnDR{B}{C}{\RR}) + \fin \;.
\end{equation}
One infers that $L_{(A)BC}^{\ONE,{\RR}}$ is determined by $2$-point functions only and finiteness  follows from the finiteness of the $2$-point functions.
This implies that the truly $3$-point CT-information is encoded in the  $L_{ABC}^{\ONE,{\RR}}$-terms.  The results of section \ref{sec:multiple} apply  straightforwardly. 
The RGE assumes the form
\begin{equation}
\label{eq:RGE2}
   ( {\cal L}_{\be} - 2 \eps)   \lnDR{A}{BC}{\RR} = -   
\zamR{A}{BC}{\RR} \;,  
\end{equation}
where 
\begin{equation}
\label{eq:chiABC}
\zamR{A}{BC}{\MS}  =2 \left(1+\frac{1}{2}(\xi^{A}+\xi^{B}+\xi^{C})+\frac{1}{2} \xi^{Q}  
 g^Q \partial_Q  \right) \rnDI{A}{BC}{1} \;, \quad \lnDR{A}{BC}{\MS} = \sum_{n \geq 1} \frac{\rnDI{A}{BC}{n} }{\eps^n} \;,
\end{equation}
and ${\cal L}_{\be}$  denotes the Lie derivative, acting on a $3$-tensor,  as in the previous section
\begin{equation}
{\cal L}_{\be} \lnDR{A}{BC}{\RR}  =   \partial_A \hat{\be}^D  \lnDR{D}{BC}{\RR}+\partial_B \hat{\be}^D     \lnDR{A}{DC}{\RR} +\partial_C \hat{\be}^D  \lnDR{A}{BD}{\RR} +  \hat{\be}^D \partial_D   \lnDR{A}{BC}{\RR} \;,
\end{equation}
where $\partial_A \hat{\be}^D =  \hat{\ga}_{A}^{\phantom{A} D}$ is the anomalous dimension matrix \eqref{eq:galog}. The finiteness of $\lnDR{A}{BC}{\RR}$ and 
$\lnDR{\theta}{\theta\theta}{\RR}$ follows from the same arguments as in section \ref{sec:multiple} and we caution the reader that the couplings $g^{A,B,C}$ refer to couplings governing  the dynamics 
as otherwise the refined conditions apply.

All operators being the same is an interesting special case leading to the expected reduction in 
the kinematics. 
\begin{eqnarray}
\label{eq:GABC2}
\Gamma_{ABC}(p_A^2,p_B^2,p_C^2)    = 
 \cwi{(A)}{BC} (p_Q^2)  P_3   +\cwi{A}{BC} (p_Q^2)  \lambda_3  \;,
 \end{eqnarray}
 with complete symmetry in $A$, $B$ and $C$, where
\begin{alignat}{2}
& \lambda_3&\;\;=\;& p_A^4+p_B^4+p_C^4-2(p_A^2 p_B^2 + p_A^2 p_C^2+ p_B^2 p_C^2)  \;, \nonumber \\[0.1cm]
& P_3  &\;\;=\;&  p_A^4  + p_B^4+p_C^4 \;,
\end{alignat}
are the important kinematic  K\"all\'en-function and the $2$-point function structure respectively. Finiteness of  the three-point function of TEMT 
follows from  similar arguments as in section \ref{sec:multiple}. 

\section{Summary and Conclusions}
\label{sec:conclusions}

In this work we have investigated the logarithmically  divergent  CTs 
of $2$- and $3$-point  functions. Using the $d$-dimensional 
renormalisation group, convergence criteria 
have been given for asymptotically free   and safe UV-FPs in equations 
\eqref{eq:convAF} and \eqref {eq:convAS} respectively.  
This is followed by an explicit discussion of the $\vev{G^2G^2}$- and 
$\vev{\TEMTO \TEMTO}$-correlators 
and the $\vev{\bar q q \bar q q}$-correlators in QCD-like theories in section 
 \ref{sec:QCD-like} and appendix \ref{app:mass}.
By taking into account all orders the former two 
were shown to be finite but divergent at fixed order  perturbation theory  implying that the $\eps \to 0$ and the perturbation expansion do not commute in general. 
Hence fixed order results can give the wrong indication about convergence. 
Finiteness implies that the bare correlators satisfy unsubtracted dispersion relations 
and are in principle observable since they are finite and scale indepdendent  cf. sections  
\ref{sec:disp} and  \ref{sec:physical}. An application of the latter is given by 
the flow of the $\Box R$-anomaly which is related to the zero momentum limit  of the $p^4$-structure  of the TEMT correlation  function \cite{PZprep}.
Using a recent computation and the quantum action principle the $R^2$-anomaly was 
extended in the $\MS$-scheme to  NNLO (${\cal O}(\alsp{5})$) in section \ref{app:R2-anomaly}.
Generalisations of the finiteness conditions  to several couplings, assuming a conformal UV-FP, 
and $3$-point functions were presented in sections \ref{sec:multiple}  
and \ref{sec:3ptMom} respectively.

In what follows we discuss specific applications of the finiteness and the possibility of adding finite CTs 
to the UV action. A crucial point is that correlation functions which are finite with operators without anomalous 
scaling are RG-scale independent and therefore can serve as observables. 
In this view the finiteness of the $\vev{\TEMTO \dots \TEMTO}$-correlators are our most important results. 
From it follows that the $R^2$-anomaly \eqref{eq:VEVTEMT} is always proportional to $\be$-functions 
of the couplings \eqref{eq:B}. 
Furthermore the finiteness serves as the basis for showing 
that the difference of the UV and IR  $\Box R$-anomaly \eqref{eq:VEVTEMT}  is 
flow-independent\footnote{Finiteness is also a crucial ingredient to the $a$-theorem 
 as it allows to establish positivity via an unsubtracted dispersion relation  \cite{KS11,LPR12}.}  
as well as the existence of a scheme for which the $R^2$-anomaly vanishes along the entire RG-flow 
\cite{PZprep}.

Hence generally CTs are meaningful only when they are related to observables. 
Otherwise they are arbitrary (i.e.~scheme-dependent) 
since one can not forbid adding local terms to the UV action in general. 
It is though our opinion that the local terms are not arbitrary to the point that 
they ought  RG-scale independent since  the bare partition function is scale independent.   
Concerning observables one may distinguish the following two cases. 
Either the CTs drop out in the observable(s) or not. 
In the latter case  they might either be fixed by 
some other principle or they need to be determined experimentally. 
The first two  examples mentioned in the previous paragraph are 
of the the first type mentioned.  
The bare contact term  vanishes when taking the scale derivative \eqref{eq:B} 
or in the difference of the UV and IR $\Box R$-value. 
The third example of the  $R^2$-anomaly concerns a scheme-dependent question 
and corresponds to a reorganisation of terms in the trace of the energy momentum tensor. 
Let us mention two well-known examples where (divergent) CTs can be handled by symmetry.
For the vacuum polarisation, the correlation function of two electromagnetic currents, 
the dispersion integral needs to be subtracted once but the value of the subtraction is fixed by  gauge 
symmetry (zero photon mass).  The chiral anomaly, which can be regarded as coming from a divergent contact term, can not be removed by a local term while maintaining  gauge symmetry. 
Another example where this ambiguity is settled  by a symmetry, namely chiral symmetry, is 
the  low energy constant $L_{10}$ from chiral perturbation theory. 
The quantity $L_{10}$ is  related to a dispersion relation 
of the correlation function of left and right-handed octet currents ($L_{10} \sim \Pi_{\textrm{LR}}(q^2=0)$ with 
pion-pole subtracted). 
Chiral symmetry in the UV forbids us to add a contact term to  $\Pi_{\textrm{LR}}$ since 
the latter is sensitive to chiral symmetry breaking.

\appendix
\numberwithin{equation}{section}

\subsection*{Acknowledgements}

In the course of writing these papers we have benefited from discussions 
with Richard Ball, Peter Boyle, Luigi Del Debbio, Einan Gardi, 
Tony Kennedy, Zohar Komargodski, Hugh Osborn,  Guido Martinelli, 
Agostino Patella, Roberto Pellegrini, Antonin Portelli, 
Misha Shifman, Kostas Skenderis, 
Andreas Stergiou  
and especially Graham Shore. We are grateful to Ben Pullin and Saad  Nabeebaccus for thorough proofreading of the manuscript. 
The authors would like to express their gratitude  to the Mainz Institute for Theoretical Physics (MITP), the Università di Napoli Federico II and INFN for its hospitality and its partial support during the completion of this work. VP thanks Weizmann for hospitality during completing final stages of this work.

\subsection*{Addendum on notational changes in v2}

As compared to our v1 on the arXiv we believe to have improved the notation by 
adapting $2 \zamDR{Q}{\RR}  \to \zamDR{Q}{\RR} $ (defined in \eqref{eq:RGE1})
 $ \de \cwiD{x} \to \lD{x}$ (e.g. \eqref{eq:subtract})  where the latter stands for local  and we changed $\cwiD{s} \to \cwiD{\theta}$ when referring 
to the operator-part (cf. \eqref{eq:TEMT} for a  more generic decomposition) $\TEMTO = \be^A [O_A] $ of the trace of the energy momentum tensor. In addition we have indicated the scheme 
with labels in more consequence since scheme-independence is a key feature for orientation and consistency.

\section{Some additional formulae for  relevant to leading Logarithm }
\label{app:A}

\subsection{Form of leading Logarithms of $\cwiD{\gsym}$}
\label{app:LL}

The leading terms in the bare correlation function take the form 
\begin{equation}
\int d^d x e^{i p \cdot x} \matel{0}{O_g(x) O_g(0) }{0}_{\textrm{LL}} = k \sum_{n \geq 0} \frac{ ( \be_0 \als_0)^{n-1} }{\epsilon^n}
\left(  \frac{ \mu^2}{p^2} \right)^{n\epsilon}  \;,
\end{equation}
with $k$ being a constant which is immaterial for the argument.  Upon renormalising the operator $[O_g] =Z_{G^2} O_g $ 
and the coupling $\als_0 =   \als Z_{\als}$ with $Z_{G^2} =   Z_{\als} $ in the LL approximation one finds 
\begin{equation}
\int d^d x e^{i p \cdot x} \matel{0}{[O_g(x)][ O_g(0)] }{0}_{\textrm{LL}}= k \sum_{n \geq 0} \frac{ f_{n}( \be_0 \als)^{n-1} }{\epsilon^n}  \;,
\end{equation}
where 
\begin{equation}
f_n= \sum_{j=0}^{n-1}(-1)^{j} \left(  \frac{ \mu^2}{p^2} \right)^{(n-j)\epsilon} 
\binom{n}{n-j} \;.
\end{equation}
This sum evaluates to
\begin{align} 
\label{eq:nice}
\notag
f_n &=  \sum_{j=0}^{n-1} (-1)^{j}\binom{n}{n-j}  + \frac{\epsilon^n}{n!} \ln^{n}{\left(  \frac{ \mu^2}{p^2} \right)} \sum_{j=0}^{n} (-1)^{j} (n-j)^{n} \binom{n}{n-j}  \\ 
&= (-1)^{n+1}+ \epsilon^n  \ln^{n}{\left(  \frac{ \mu^2}{p^2} \right)} \;,
\end{align}
confirming the rule $\epsilon^{-n}\leftrightarrow     -  \ln^n(p^2/\mu^{2}) $  used in section \ref{sec:disp}.

Note that non-local divergent terms in  \eqref{eq:nice} are avoided since the sum, somewhat magically, 
\begin{equation}
 \sum_{j=0}^{n-1} (-1)^{j} (n-j)^{l} \binom{n}{n-j}  = 0 \;, \quad 0<l<n \;,
\end{equation}
only contributes for $l = 0$ and $l = n$. We note that such non-local terms could not be eliminated by 
local counterterm in perturbation theory.

\subsection{The leading poles of the counterterm $\lDR{Q}{\MS}$ for $\xi_Q \neq 0$}
\label{app:detail} 

The leading poles of the counterterm $\lDR{Q}{\MS}$ \eqref{eq:LQu}  for an AF theory in the 
$\MS$-scheme for $\xi_Q \neq 0$ is given by 
 \begin{eqnarray}
\label{eq:hyper}
\lDR{Q}{\MS}(\mu)  &\simeq&  2 (1+ \xi_Q) \rDI{Q}{1,0}  \int_{0}^{\infty}  e^{-2 (1+ \xi_Q)\eps t} \eps^{-\frac{\gaZ{Q}}{\be_0}}  (\eps+\be_0 \als (1-e^{-2\eps t})^{\frac{\gaZ{Q}}{\be_0}}) dt 
\nonumber  \\[0.1cm]
&=& 
   \rDI{Q}{1,0} \Big( \frac{(1+ \frac{\als \be_0}{\eps})^{1+\frac{\gaZ{Q}}{\be_0}}(-1-\frac{\eps}{\be_0 \als})^{\xi_Q}\Gamma(1+\xi_Q)\Gamma(-\frac{\gaZ{Q}}{\be_0}- \xi_Q)}{ \als (\be_0(1+\xi_Q)+\gaZ{Q})\Gamma(-\frac{\gaZ{Q}}{\be_0})}
\nonumber  \\[0.1cm]
& &+ \frac{2 \be_0 (-\eps)^{-\frac{\gaZ{Q}}{\be_0}}{}_2F_1\left[-\frac{\gaZ{Q}}{\be_0},-1-\frac{\gaZ{Q}}{\be_0}+\xi_Q,-\frac{\gaZ{Q}}{\be_0}+\xi,\frac{a \be_0+\eps}{a \be_0}\right]}{\eps (\be_0(1+\xi_Q)+\gaZ{Q})}  \Big)\;,
 \end{eqnarray}
where $\xi_Q $ is the difference of the $d$-dimensional and four dimensional 
anomalous dimension $\hat{\ga}_Q = \ga_Q - \xi_Q \eps$. For $\xi_Q \to 0$ the formula 
simplifies considerably and is given in section \ref{sec:AF} in \eqref{eq:LQresult}.

\subsection{Explicit evaluation of the Dispersion Integral}
\label{app:check}
As a check the integral \eqref{eq:xh} is integrated explicitly. This is best done by changing variables to 
 $s = \mu^2 e^y$ which results in an integral (recall $P^2 = - p^2$)
\begin{equation}
 \hat{x}(P^2) = \int_{- \infty}^\infty dy \frac{dy e^y}{ e^y - P^2/\mu^2}\frac{1}{(1 + \als \be_0 y)^2 + (a \beta_0 \pi)^2} \;,
\end{equation}
with a poles at $y_{\pm} =  - 1/(\als \be_0) \pm i \pi$ and a series of poles $y_{n_\pm} = \ln (-P^2/\mu^2) \pm i \pi (2n+1)$ for $n \geq 0$. The integration contour can, for example, be closed in the upper half plane.  
The $y_{+} $ pole result in the pole term in \eqref{eq:xdisper} and the series of poles 
$y_{n_+} = \ln (-P^2/\mu^2) + i \pi (2n+1)$ for $n \geq 0$ leads to a series
\begin{eqnarray}
 \hat{x}(P^2) &=&  \frac{1}{ 1 -      P^2/P_0^2} - ( 2 \pi i)\sum_{n \geq 0} \frac{1} {1+ \als \be_0 ( \ln  (-P^2/\mu^2) + i \pi( 2n+1))^2+    (a \beta_0 \pi)^2}  
   \nonumber \\ 
  &=& \frac{1}{ 1 -      P^2/P_0^2}  + x(P^2)  \;,
\end{eqnarray}
 which can be resumed into an analytic form. The final result is consistent with Eq.\eqref{eq:xdisper} which was 
the aim of this appendix.

\section{Quark Current Correlators}
\label{app:quark}

\subsection{The $\vev{\bar{q}q\bar{q}q}$-correlator in QCD-like Gauge Theories}
\label{app:mass} 

Finally we consider the bifermion scalar operator 
\begin{equation}
[O_M] = [\bar q q]\;, \quad   \kappa_M = m     \;,
\end{equation}
for which $m [\bar q q] = m_0 \bar q q$ is an RG-invariant. The parameter $m$  
does not enter the dynamics and is regarded as a source term only.  
The relevant input to  criteria \eqref{eq:convAF} is given by 
$\gaZ{M}$,  $\zam{M}{M}$ and  $\be_0$.
The leading order of the mass anomalous dimension is given by ($\ga_{\bar q q} =  \ga_M = - \ga_m$ and $\hat{\ga}_m = \ga_m$ since $\bar q q$ is a kinetic operator)
\begin{equation}
  \ga_M = \gaZ{M} \als + {\cal O}(\alsp{2}) \;, \quad  \gaZ{M}  = - 6 C_F  \;,
\end{equation}
where  $\rnDI{M}{M}{1}(\als) =\rnDI{M}{M}{1,0}  + {\cal O}(\als)$, $\be_0$  and $C_F$  are given in \eqref{eq:bei} and \eqref{eq:SUNc}. 
With $\zamR{M}{M}{\RR} = {\cal O}(\alsp{0})$ (i.e. $n_{MM}=0$) condition \eqref{eq:convAF} reads
\begin{equation}
\label{eq:m-criteria}
-\frac{\gaZ{M}}{\be_0} \Big|_{SU(N_c)} = \frac{3 (N_c^2-1)/( N_c)}{11/3 N_c - 2/3 N_f}  > 1
 \quad \Leftrightarrow \quad  \lD{M} = \fin 
 \;.
\end{equation}
This criteria is satisfied for 
$N_f >  (9+2 N_c^2)/(2 N_c) $ which for $N_c = 3$ leads to 
convergence for   $N_f > 4.5$.

The leading pole contribution \eqref{eq:LQresult} is given by
\begin{equation} 
\label{eq:LmLL}
\lDR{M}{\MS}|_{\textrm{LL}} =   \rnDI{M}{M}{1,0}  \frac{(1+ \frac{\als \be_0}{\eps})^{1+\frac{\gaZ{M}}{\be_0}}-1}{ \als (\be_0+\gaZ{M})} 
 \stackrel{\mbox{\eqref{eq:m-criteria}} }{\longrightarrow} \lDfR{M}{\MS}|_{\textrm{LL}} =   
- \frac{\rnDI{M}{M}{1,0}}{ \als (\be_0 + \gaZ{M} )}  \;.
\end{equation}
where we have assumed \eqref{eq:m-criteria} to obeyed.
For QCD with three massless flavours $N_f =3$ and $N_c = 3$ the expression is divergent.
Presumably this means that the constant part of the $\Gamma_{MM}$-correlator is not directly related to a physical quantity. 
Expanding 
in $\als$ one obtains \eqref{eq:LQresult}
\begin{equation}
\lDR{M}{\MS}|_{\textrm{LL}} = \rnDI{M}{M}{1,0}  \left( \frac{1}{\eps}+\frac{\gaZ{M} \als}{2 \eps^2}+\frac{(-\be_0 
\gaZ{M}+(\gaZ{M})^2)\alsp{2}}{6  \eps^3} + {\cal O}(\alsp{3}) \right) \;,
\end{equation}
from where the leading poles in  \cite{Chetyrkin:1996sr,Narison:1989aq} are recovered. 
For the sake illustration let us quote the LL result,  obtained by replacing $\frac{1}{\eps} \to - \ln {\left(  \frac{ p^2}{\mu^2} \right)}$, 
\begin{eqnarray}
\Gamma_{MM}^\MS|_{LL}(p^2) &=& \int d^4 x e^{i p \cdot x} \matel{0}{[\bar{q}q(x)][ \bar{q}q(0)] }{0}_{\textrm{LL}} \nonumber \\[0.1cm]
&=& 
- p^2 \rnDI{M}{M}{1,0} \frac{(1 + \als \be_0 \ln {\left(  \frac{ p^2}{\mu^2} \right)})^{1+\frac{\gaZ{M}}{\beta_0}}-1}{ \als (\be_0 + \gaZ{M})} + \dots \;,
\end{eqnarray}
where the dots stand for condensate contributions.
Expanding in $\als \ln {\left(  \frac{p^2}{ \mu^2} \right)}$ the  ${\cal O}(\alsp{3})$-LL expression matches the result in 
\cite{Chetyrkin:1996sr}.

Following section \ref{sec:physical} we explicitly demonstrate at LL that the bare correlator, multiplied by $\kappa_M^2 = m^2(\mu)$, is $\mu$-independent in the following sense
\begin{eqnarray}
\label{eq:mgeneral}
m^2(\mu) \Gamma_{MM}(p^2,\mu)  &=&  \mu_0^4 f( \als(\mu^2/\mu_0^2),m/\mu_0,p^2/\mu_0^2)  \nonumber \\[0.1cm]
&=& p^2 m^2(p^2) F( \als(p^2/\mu_0^2) ) \  \;,
\end{eqnarray}
 and $\mu_0$ being an arbitrary reference scale.
 First we note that the renormalised correlator 
\begin{equation}
\label{eq:mfunction}
m^2(\mu) \Gamma^\MS_{MM}|_{\textrm{LL}}(p^2) = p^2 \left(- \rnDI{M}{M}{1,0}  \frac{m^2(p^2)}{ \als(p^2) (\be_0 + \gaZ{M})}+  \rnDI{M}{M}{1,0} \frac{m^2(\mu)}{ \als(\mu) (\be_0 + \gaZ{M})} \right) \;,
\end{equation}
splits into a $\mu$-independent non-local and a $\mu$-dependent local term. If we now restrict to the convergent case satisfying \eqref{eq:m-criteria}, then 
the second term is equal to \eqref{eq:LmLL} and in the $\eps \to 0$ limit
\begin{eqnarray}
m^2(\mu) \Gamma_{MM}(p^2) &=& m^2(\mu) \Gamma^{\MS}_{MM}(p^2) + \lDfR{M}{\MS} \nonumber 
\\[0.1cm]
&\stackrel{\textrm{LL}}{=}&   - p^2  \rnDI{M}{M}{1,0}  \frac{m^2(p^2)}{ \als(p^2) (\be_0 + \gaZ{M})}   \;,
\end{eqnarray}
which satisfies \eqref{eq:mgeneral} in  analogy with \eqref{eq:CWLL}.

\subsection{The $\vev{ J_\mu^5J_\nu^5} $-correlator}
\label{app:55} 

The axial current $2$-point function in an AF-theory has been studied by Shore \cite{S91} 
and is worthwhile to be captured  language of this paper.
The correlator decomposes into 
\begin{equation}
\int d^4 x ^{i x \cdot p} \vev{ J_\mu^5(x) J_\nu^5(0)} = (\de_{\mu \nu} p^2 - p_\mu p_\nu) \CW^{\ONE, T}_{J_5 J_5}(p^2) 
+ p_\mu p_\nu  \CW^{\ONE, L}_{J_5 J_5}(p^2) \;,
\end{equation}
a transversal (T) and a longitudinal (L) part.   
 Since $\ga_{J_5,0} = 0$,  the  criteria   \eqref{eq:convAF} implies convergence for $n^{T,L}_{J_5 J_5} > 1$ where 
$\chi^{T,L}_{J_5 J_5} \sim \als^{n^{T,L}_{J_5 J_5}}$ is defined in analogy to \eqref{eq:RGE1}.  
In the case of massless fermion considered here the axial current correlation function is identical 
to the vector current correlation function (vacuum polarisation). Hence 
the important ingredient to the analysis 
is the conservation of the vector current which implies that the transverse part 
contributes at LO $n^{T}_{J_5 J_5} = 0$   
and further implies that $n^{L}_{J_5 J_5} > 0$. 
Thus the well-known LO divergent contact term of the vacuum polarisation is not resummed 
to a finite expression. 
Yet in the longitudinal part the chiral anomaly itself contributes at
 NNLO, with $n^{L}_{J_5 J_5} = 2$, which then implies convergence 
and a scaling of the type 
$ \CW^{\ONE, L}_{J_5 J_5}(p^2) \sim \als $ in analogy to the TEMT-correlator \eqref{eq:CWLL}. 
This result   
is consistent with eq.~(6.36) of Shore's work \cite{S91}.

\section{Conventions for $\be$-function}
\label{app:beta}

In this work the bare $\be$-function $\hat{\be}$ of DR is defined as
\begin{equation}
\label{eq:belog}
\hat{\beta}= \frac{d \ln g}{d \ln \mu}= \frac{(d-4)}{2} + \beta  =  - \eps + \be \;.
\end{equation}
We draw the reader's attention to the fact that the logarithmic $\be$-function \eqref{eq:belog}
is used throughout in order to keep the formulae more compact. Explicitly 
\begin{equation}
\label{eq:bei}
\beta= - \beta_0 \als - \beta_1 \alsp{2} - \beta_2 \alsp{3} - \beta_3 \alsp{4}+  \dots \;, \quad \als = \frac{\al_s}{4 \pi} =  \frac{g^2}{(4 \pi)^2} \;,
\end{equation}
where $\be_{0-3}$   in $\overline{\MS}$-scheme can be found in Ref.~\cite{CZAKON}.
The first two coefficients, which are universal in mass-independent schemes, read 
$$\beta_0 = ( \frac{11}{3} C_A- \frac{4}{3} N_F T_F) \;, \quad  \beta_1 =  (\frac{34}{3} C_A^2- \frac{20}{3} N_c  N_F T_F - 4 C_F T_F  N_F)   \;, $$  
 where  $C_F$, $C_A$ are the quadratic Casimir operators of the fundamental (quark) and adjoint (gluons) 
 representations, $N_F$ the number of quarks and $ \tr[T^a T^b] = T_F \delta^{ab}$ is a 
 Lie algebra  normalisation factor  of the fundamental representation. For $SU(N_c)$ these factors are given by
 \begin{equation}
\label{eq:SUNc}
  C_A = N_c \;, \quad  C_F = \frac{N_c^2-1}{2 N_c}  \;, \quad T_F = \frac{1}{2}  \;.
 \end{equation}

\bibliographystyle{utphys}
\bibliography{input2}

\end{document}